\documentclass[a4paper,11pt,fleqn]{article}

\usepackage[utf8]{inputenc}

\usepackage{amsmath}
\usepackage{mathrsfs}
\usepackage{amssymb} 
\usepackage{amsfonts}

\usepackage{cite}

\usepackage[hang,small,bf]{caption}

\usepackage{graphicx}

\usepackage{a4}
\usepackage[T1]{fontenc}

\usepackage{geometry}
\usepackage{float}

\usepackage{makeidx}

\usepackage{vector}
\usepackage{slashed}
\usepackage{color}

\usepackage{bm}

\usepackage{pstricks}

\usepackage{feynmf}

\usepackage{subfigure}

\usepackage{float}
\restylefloat{figure}
\usepackage{ae}
\usepackage{times}
\usepackage[colorlinks=true,linkcolor=black,citecolor=black,bookmarks=true,bookmarksopen=true,pdfpagemode=None,pdfstartview=FitH]{hyperref}
\hypersetup{colorlinks=false}

\usepackage{graphicx,epsfig}

\newcommand{\scalefac}{a}
\newcommand{\hubblerate}{H}

\definecolor{grey}{rgb}{0.5,0.5,0.5}

\newcommand{\eqn}{eqn.\,}

\newcommand{\dend}{{\, .}}

\newcommand{\kend}{{\, ,}}

\newcommand{\dpend}{\, .}

\newcommand{\ifeqthenelse}[4]{\edef\tempa{#1}\def\tempb{#2}\ifx\tempa\tempb {#3} \else {#4}\fi}

\newcommand{\formula}[2]{$#1\ifeqthenelse{#2}{}{}{\label{eqn:#2}}$}
\newcommand{\bigformula}[2]{\begin{equation}#1\ifeqthenelse{#2}{}{}{\label{eqn:#2}} \end{equation}}
\newcommand{\bigformularray}[2]{\begin{eqnarray}#1\ifeqthenelse{#2}{}{}{\label{eqn:#2}}\end{eqnarray}}

\newcommand{\eqnref}[1]{(\ref{eqn:#1})}

\newcommand{\salign}{&&\hspace{-6mm}}

\newcommand{\plottitle}{}

\newcommand{\fig}{fig.\,}
\newcommand{\figref}[1]{\ref{fig:#1}}

\newcommand{\app}{appendix\,}
\newcommand{\appref}[1]{\ref{app:#1}}

\newcommand{\sect}{section\,}
\newcommand{\secref}[1]{\ref{sec:#1}}

\newcommand{\dkn}[1]{#1}

\newcommand{\todo}[1]{}

\newcommand{\order}[1]{\mathcal{O}(#1)}

\newcommand{\floorbr}[1]{\lfloor #1\rfloor}

\newcommand{\const}{\mathrm{const}}

\newcommand{\sgn}{\mathrm{sgn}}

\newcommand{\ramp}[1]{R#1}

\newcommand{\abs}[1]{\left|#1\right|}

\newcommand{\heaviside}[1]{\Theta\ifeqthenelse{#1}{}{}{\left(#1\right)}}

\newcommand{\Min}[1]{\min\left(#1\right)}

\newcommand{\Max}[1]{\max\left(#1\right)}

\newcommand{\sbessel}[2]{j_{#1}(#2)}

\newcommand{\partd}[2]{\frac{\partial #1}{\partial #2}}

\newcommand{\diff}[1]{{d}#1}

\newcommand{\Dnm}{D^{nm}}
\newcommand{\Dnmind}[2]{D^{#1,#2}}

\newcommand{\coeffa}[2]{a_{#1,#2}}

\newcommand{\termc}[1]{\left( #1 \right)}
\newcommand{\termtheta}{\Theta\left( k,p,q,r \right)}
\newcommand{\termprefac}[1]{#1\termtheta}
\newcommand{\funcU}[5]{U_{#1}\left(#2,#3,#4,#5\right)}

\newcommand{\intIn}[9]{I(#1;#2,#3,#4,#5;#6,#7,#8,#9)}
\newcommand{\intIthree}[6]{I(#1,#2,#3;#4,#5,#6)}

\newcommand{\wignerthreej}[6]{
\begin{pmatrix} 
 #1 & #2 & #3 \\ 
 #4 & #5 & #6 
\end{pmatrix}
}

\newcommand{\wignersixj}[6]{
\left\{
\begin{matrix} 
 #1 & #2 & #3 \\ 
 #4 & #5 & #6 
\end{matrix}
\right\}
}

\newcommand{\me}{\mathcal{M}}

\newcommand{\dif}{\,d}

\newcommand{\bvk}{\bvec{k}}
\newcommand{\bvp}{\bvec{p}}
\newcommand{\bvq}{\bvec{q}}
\newcommand{\bvr}{\bvec{r}}
\newcommand{\bvl}{\bm{\lambda}}

\newcommand{\meavsqu}{\left|\me\right|^2}

\newcommand{\ctkq}{\cos\left(\theta_{kq}\right)}
\newcommand{\ctkr}{\cos\left(\theta_{kr}\right)}
\newcommand{\cttwovec}[2]{\cos\left(\theta_{\bvec{#1}\bvec{#2}}\right)}
\newcommand{\ctonevec}[1]{\cos\theta_{\bvec{#1}}}

\newcommand{\stonevec}[1]{\sin\theta_{\bvec{#1}}}
\newcommand{\ctq}{\cos\theta_{\bvq}}
\newcommand{\ctr}{\cos\theta_{\bvr}}
\newcommand{\ctl}{\cos\theta_{\bvl}}
\newcommand{\stq}{\sin\theta_{\bvq}}
\newcommand{\str}{\sin\theta_{\bvr}}
\newcommand{\stl}{\sin\theta_{\bvl}}

\newcommand{\indl}{n}
\newcommand{\indk}{\alpha}
\newcommand{\vecx}{\bvec{x}}

\newcommand{\veceta}{\bvec{\eta}}
\newcommand{\vecetaunit}{{\mathbf{\hat\eta}}}
\newcommand{\veckunit}{\hat{\bvec{k}}}
\newcommand{\vecxunit}{\hat{\bvec{x}}}
\newcommand{\vecxabs}{x}

\newcommand{\vecetaabs}{\eta}
\newcommand{\laplace}{\Delta}

\begin{document}
\begin{titlepage}
\vspace*{10mm}
\begin{center}
{\Large\sffamily\bfseries
\mathversion{bold}

Solving the Homogeneous Boltzmann Equation with Arbitrary Scattering Kernel
\mathversion{normal}}
\\[13mm]
{\large
A.~Hohenegger\footnote{E-mail: \texttt{Andreas.Hohenegger@mpi-hd.mpg.de}}}
\\[5mm]
{\small \textit{
Max-Planck-Institut f\"{u}r Kernphysik\\ 
Postfach 10 39 80, 69029 Heidelberg, Germany
}}
\vspace*{1.0cm}
\end{center}
\normalsize

\begin{abstract}
\noindent With applications in astroparticle physics in mind, we generalize a method for the solution of the nonlinear, space homogeneous Boltzmann equation with isotropic distribution function to arbitrary matrix elements. The method is based on the expansion of the matrix element in terms of two cosines of the ``scattering angles''. The scattering functions used by previous authors in particle physics for matrix elements in Fermi-approximation are retrieved as lowest order results in this expansion. The method is designed for the unified treatment of reactive mixtures of particles obeying different scattering laws, including the quantum statistical terms for blocking or stimulated emission, in possibly large networks of Boltzmann equations. Although our notation is the relativistic one, as it is used in astroparticle physics, the results can also be applied in the classical case.
\end{abstract}

\end{titlepage}

\setcounter{footnote}{0}

\newpage

\section{Introduction}
Non-equilibrium processes in astroparticle physics, such as Big Bang nucleosynthesis (BBN), neutrino decoupling and more speculative ones, like baryogenesis through leptogenesis or the freeze-out of hypothetical relic particles, \cite{Hannestad:1995rs,Gnedin:1998,Dodelson:1992,Kolb:1990,Fukugita:1986,Buchmuller:2004,Serpico:2004gx} are usually computed through the solution of the corresponding coupled system of Boltzmann equations \cite{Bernstein:1988,Groot:1980,Mischler:2003,Cercignani:2002,Liboff:2003} describing the time evolution of the one-particle distribution functions $f^i(t,k)$. Typically, in cosmology, it is anticipated that the relevant particle distributions are in exact kinetic equilibrium and of Maxwell-Boltzmann type. These assumptions, together with others, allow the Boltzmann equation to be linearized and integrated, which leads to coupled sets of chemical rate equations (mostly themselves dubbed Boltzmann equations in this context). This procedure drastically simplifies the numerical computation of the particle abundances, such that even the approximate solution of very large networks of Boltzmann equations, as in the case of BBN, becomes possible. However, in doing so, one looses the spectral information contained in the definition of the distribution functions and other fundamental properties of the Boltzmann equation are neglected as well. It is well-known that the solution of the full Boltzmann equations can lead to relevant corrections to the equilibrium results in several cases \cite{Dolgov:1997mb,Hannestad:1999fj,Basboll:2006yx}. In the era of precision cosmology the inclusion of such non-equilibrium effects gains in importance. Regarding the use of classical kinetic theory for the description of phenomena in the (very) early universe there are concerns, originating in the belief that these calculations should be performed in the framework of non-equilibrium quantum field theory. These concerns are supported by recent results, revealing differences between the two approaches for simple toy models, at least in extreme non-equilibrium situations, see e.g. \cite{Lindner:2005,Berges:2002wr}. However, it seems natural to attempt to include quantum effects in modified effective kinetic equations. Boltzmann equations will continue to play an important role in cosmology at least at the relatively low energies of neutrino decoupling or nucleosynthesis, where the standard calculations give already quite good results.

In general, a network of Boltzmann equations can be written as:
\bigformula{L[f^i]=\sum_{l} C^{il}[f^1,\ldots ,f^i ,\ldots ,f^N]\kend}{system_of_boltzmann_equation}
where there is one equation for each of the $N$ participating particle species ($i=1\ldots N$) and one collision term $C^{il}$ for each interaction with particles of the same and of other species. $L$ denotes the Liouville operator, most commonly in Minkowski space-time,
\bigformula{L[f^i](k)=\partd{f^i(t,k)}{t}\kend}{rel_liouville_uncurved}
or in Robertson-Walker space-time,
\bigformula{L[f^i](k)=\partd{f^i(t,k)}{t}-\hubblerate k\partd{f^i(t,k)}{k}\kend}{rel_liouville}
with Hubble rate $\hubblerate ={\dot{a}}/{a}$. By writing the collision integrals as $C^{il}[f^1\ldots f^i\ldots f^N]$, we have formally taken the possibility of multi-particle scattering processes into account. Usually only decays, inverse decays and $2-2$ scattering processes,  $a+b\leftrightarrow A+B$, are considered. For the latter ones the collision integral reads
\bigformularray{&&\hspace{-6mm}C^{al}[f^af^bf^Af^B](k)=\nonumber\\ 
&&\qquad\frac{1}{2E^a_k}\int\int\int\frac{\dif^3{p}}{(2\pi)^32E^b_p} \frac{\dif^3{q}}{(2\pi)^32E^A_q} \frac{\dif^3{r}}{(2\pi)^32E^B_r}  (2\pi)^4\delta^4(k+p-q-r)\meavsqu\nonumber\\ 
&&\qquad\qquad\qquad\times \left[(1-\xi^a f_k^a)(1-\xi^b f_p^b)f_q^Af_r^B-f_k^af_p^b(1-\xi^A f_q^A)(1-\xi^B f_r^B)\right]\kend\nonumber\\}{collision_integral}
where we have used the short-hand notation $f^i_k=f^i(t,k)$ and $\xi^i$ to specify the quantum statistics of particle species $i$, i.e. $\xi^i=+1$ for Fermi-Dirac, $\xi^i=-1$ for Bose-Einstein and $\xi^i=0$ for Maxwell-Boltzmann statistics. $\meavsqu$ denotes the invariant matrix element squared and averaged over initial and final spin states. Note that we take $\meavsqu$ to include possible symmetrization factors of $1/2$ for identical particles in the initial or final state. 
There is a vast number of different methods for the solution of the Boltzmann equation (mostly applied in different fields of physics), out of which only a few exploit the homogeneity and isotropy as imposed by the cosmological principle. So-called direct integration methods, where the collision term \eqnref{collision_integral} is integrated numerically, seem to be most advantageous because they are characterised by high precision This is desirable since one wants to keep track of only small deviations from equilibrium. The direct numerical solution using deterministic methods is numerically expensive, mainly because of the multiple integrals in the collision terms. Therefore, it relies on the successful reduction of the collision integral for isotropic distribution functions. 

In the present paper a technique for this reduction of the collision integral is presented which generalizes previous results in high energy and astro physics \cite{Dolgov:1997mb,Yueh:1976} for matrix elements in Fermi-approximation to (in principle) arbitrary matrix elements, relying on a series expansion of the matrix element. The resulting reduced Boltzmann equation contains only a two-fold integral over the magnitudes of the post-collisional momenta. The method is applicable to Boltzmann equations with and without quantum statistical terms and can be used independent of the dispersion relation, i.e. it can be used for massive and massless relativistic particles as well as for non-relativistic ones. The loss and gain terms can be treated collectively or independently. Thus the method represents an approach to treat reactive mixtures of all kinds of particles with different interactions, in a unified manner.

The outline is as follows: In \sect\secref{reduction_of_the_collision_integral} we show how the nine-dimensional collision integral for $2-2$ scattering processes can be reduced to a two-dimensional one, integrating out the energy and momentum conserving $\delta$-functions. (Collision integrals for decays and inverse decays can be integrated in the same way.) In doing so, a certain angular integral over the matrix element arises. In \sect\secref{a_simple_numerical_model} we establish a simple numerical model for the reduced Boltzmann equation. The integral of \sect\secref{reduction_of_the_collision_integral} is solved by expanding the matrix element in terms of the cosines of two ``scattering angles'', see \sect\secref{expansion_of_the_scattering_kernel}. In \sect\secref{numerical_integration} we derive a formula, suitable for numerical integration of the full matrix element, which we employ in the last section to demonstrate the convergence of the series torwards the exact result for a simple example. We conclude in \sect\secref{conclusion}.

\section{Reduction of the Collision Integral\label{sec:reduction_of_the_collision_integral}}
Omitting the superscripts denoting the particle species\footnote{From here on we will always use the momenta $k$, $p$, $q$ and $r$ in connection with only one particle species, such that it serves as a label for the species at the same time. We also use the convention $v=\abs{\bvec{v}}$ if the distinction from the four-momentum is clear from the context.} in \eqn\eqnref{collision_integral} we can write the collision integral as:
\bigformula{C[f](k)=\frac{1}{2E_k}\int(2\pi)^4\delta(E_k+E_p-E_q-E_r)\delta^3(\bvk+\bvp-\bvq-\bvr)\meavsqu F[f]\prod_{\bvec{v}=\bvp,\bvq,\bvr}\frac{\dif^3{v}}{(2\pi)^32E_v}\kend}{collision_integral_numerical}
where we introduced
\bigformula{F[f]=(1-\xi^k f_k)(1-\xi^p f_p)f_qf_r-f_kf_p(1-\xi^q f_q)(1-\xi^r f_r)\dend}{func_F}
$E_v$ denotes the relativistic energy of the particles ``$v$'' on the mass shell, i.e. $E_v=\sqrt{\bvec{v}^2+m_v^2}$, with three-momentum $\bvec{v}$ and mass $m_v$.
We write the 3-dimensional $\delta$-function as the Fourier transform of unity and switch to spherical coordinates: 
\bigformula{\delta^3(\bvk+\bvp-\bvq-\bvr)=\int e^{i\bvl(\bvk+\bvp-\bvq-\bvr)}\frac{\dif^3\lambda}{(2\pi)^3}\kend}{} 
The collision term \eqnref{collision_integral_numerical} then becomes 
\bigformula{C[f](k)=\frac{1}{64\pi^3E_k}\int\delta(E_k+E_p-E_q-E_r)F[f]D(k,p,q,r)\frac{p\dif p}{E_p}\frac{q\dif q}{E_q}\frac{r\dif r}{E_r}\dend}{simplified_collision_integral} 
Here we have defined $D$ as 
\bigformularray{
D(k,p,q,r)&=& \frac{pqr}{8\pi^2}\int \dif\Omega_p\int \dif\Omega_q\int \dif\Omega_r\,\, \delta^3(\bvk+\bvp-\bvq-\bvr)\meavsqu\nonumber\\
&=&\frac{pqr}{64\pi^5}\int \lambda^2\dif\lambda\int e^{i\bvl\bvk}\dif\Omega_\lambda\int e^{i\bvl\bvp}\dif\Omega_p\int e^{-i\bvl\bvq}\dif\Omega_q\int e^{-i\bvl\bvr}\dif\Omega_r\meavsqu\dend\nonumber\\
}{D_function}
Note, that this definition renders $D(k,p,q,r)$ a dimensionless quantity. Due to the presence of the $\delta$-function we expect that the result is non-zero only if $q+r>\abs{k-p}$ and $k+p>\abs{q-r}$, because the equation $\bvk +\bvp =\bvq +\bvr$ does not have a solution otherwise, for whatever combination of the solid angles $\Omega_p$, $\Omega_q$, $\Omega_r$. Therefore, the result will be proportional to\footnote{Throughout we use the Heaviside-step-function $\heaviside{}$ in the half-maximum convention (which will become relevant later).}
\bigformularray{\termprefac{}&\equiv&\heaviside{q+r - \abs{k-p}}\,\heaviside{k+p-\abs{q-r}}\nonumber\\ & =&\heaviside{\min\left({k+p,q+r}\right)-\max\left(\abs{k-p},\abs{q-r}\right)}\dend}{heaviside_prefactor}

After computing $D(k,p,q,r)$ we can proceed with the integration of the remaining energy $\delta$-function in \eqn\eqnref{simplified_collision_integral}:
\bigformula{C\left[f\right](k)=\frac{1}{64\pi^3E_k}\int\int\heaviside{E_p-m_p} F[f] D(k,p,q,r)\frac{q\dif q}{E_q}\frac{r\dif r}{E_r}\kend}{boltzmann_equation_reduced} 
where $p=\sqrt{E_p^2-m_p^2}$ and $E_p=E_q+E_r-E_k$. The Heaviside-function prevents us from integrating over combinations of $q$ and $r$ which are kinematically forbidden. Thus we have reduced the collision integral to a two-dimensional one, suitable for numerical integration. However, all the work is now hidden in the definition of $D=D(k,p,q,r)$ which is characteristic for the scattering model, i.e. for the matrix element of the underlying theory for the scattering process under consideration. For completeness we present the analogous calculation for $C^{1\leftrightarrow 2}$ like collision integrals in \app\appref{C12_like_collision_integrals}.

The computation of $D$ is easily carried out for matrix elements squared with simple angular dependence such as the constant ($\meavsqu=\const$), for matrix elements in the Fermi approximation ($\meavsqu\propto (k\cdot p)(q\cdot r), (k\cdot p)$, including re-namings of the momenta therein) and for resonant processes in the narrow width approximation ($\meavsqu\propto\delta (s-m_X^2)$, where $m_X$ is the mass of the particle in the intermediate state).
However, in particle physics, one encounters matrix elements squared with a more complicated structure, such as products of tree-level $s$-, $t$- and $u$-channel contributions, for which the integrals $D$ are in general unknown.

\newcommand{\x}{x}
\section{A Simple Numerical Model\label{sec:a_simple_numerical_model}}

In this section we establish a simple numerical model to benefit from the reduced form of the collision integral. For simplicity we assume a single particle species undergoing $2-2$ scattering processes only. The system \eqnref{system_of_boltzmann_equation} then acquires the form 
\bigformula{L[f]=C[f]\kend}{single_boltzmann_equation}
with $C[f]$ from \eqn\eqnref{boltzmann_equation_reduced} and $L[f]$ either from \eqn\eqnref{rel_liouville} or \eqnref{rel_liouville_uncurved}.

In case the Liouville operator of the system is of the first kind with the extra term \bigformula{-Hk\partd{f^i(t,k)}{k}\kend}{} as compared to \eqn\eqnref{rel_liouville_uncurved}, which accounts for the expansion of the universe\footnote{It is this term which prevents the Maxwell-J{\"u}ttner distribution function in general, from being an exact equilibrium solution for the Boltzmann equation in Robertson-Walker space-time. Only if the interaction rate is high enough its effect on the shape of the distribution function can be neglected.}, we first introduce the transformed variables
\bigformularray{\x=M\scalefac(t)\kend\quad \tilde{k}=k\scalefac(t)\kend}{variable_transform}
with some suitable mass scale $M$ and cosmic scale factor $\scalefac(t)$.
The Liouville operator in this new coordinates has the simpler form  
\bigformula{L[f]=H\x\,\frac{\partial f(\x,\tilde{k})}{\partial \x}\dend}{rel_liouville_transformed}
In the subsequent we omit the tilde over the transformed momenta and time, but it is important to remember that, in this case, the collision integrals need to be expressed in terms of the transformed variables as well.

Now we divide the physical relevant part of the positive real axis of momenta, i.e. we consider only momenta up to a maximum of $k_{\text{max}}$, into a set of $M$ disjoint (not necessarily equidistant) intervals $\Delta k_i$ and choose a $k_i$ for each interval with $k_i\in \Delta k_i$ ($i=1\ldots M$). 

Then we make the approximation 
\bigformula{\int_{\Delta k_i}f(t,k)\dif k\simeq f(t,k_i)\Delta k_i\equiv f_i\Delta k_i\dend}{} 

\newcommand{\smallDk}{\Delta  k}
By integrating the left-hand side of  \eqnref{single_boltzmann_equation} over the interval $\Delta k_l$ we obtain 
\bigformula{L_l=\int_{\Delta k_l}\frac{\partial f_k}{\partial t}\dif k\simeq\frac{\partial f_l}{\partial t}\Delta k_l\kend\quad\text{or}\quad L_l=\int_{\Delta k_l}H\x\frac{\partial f_k}{\partial \x}\dif k\simeq H\x\frac{\partial f_l}{\partial \x}\Delta k_l\kend}{discrete_liouvil_operator}

For the right-hand side, we find
\bigformularray{&&\hspace{-6mm}C_l=\frac{1}{64\pi^3E_{k_l}}\sum_{\underset{E_p\geq m_p, p\leq k_{max}}{i,j}}^M\left[(1-\xi f_l)(1-\xi f_p)f_if_j-f_lf_p(1-\xi f_i)(1-\xi f_j)\right]\nonumber\\
&&\hspace{45mm}\times D(k_l,p,k_i,k_j)\frac{k_i\smallDk_i}{E_{k_i}}\frac{k_j\smallDk_j}{E_{k_j}}\kend}{discrete_collision_integral}
where $p=\sqrt{(E_{k_i}+E_{k_j}-E_{k_l})^2-m_p^2}$. 

This way, we turned the reduced Boltzmann equation into a coupled set of $M$ ordinary differential equations for the discrete variables $f_l$: 
\bigformularray{L_l =C_l ,\qquad (l=1\ldots M)\dpend}{discrete_boltzmann_equation}

Due to the finite momentum cutoff $k_{max}$, this method is not conservative, i.e. energy and total particle number are not conserved inherently. In order to make the method energy and particle number conserving, the cutoff has to be chosen high enough.

The number of equations, or grid points $M$, depends on the specific problem and the required accuracy. In any case \eqn\eqnref{discrete_boltzmann_equation} will represent a large system of differential equations. The amount of numerical work for the evaluation of the collision integral is of order $\order{M^3}$.

$D$ can in principle be computed numerically, see \sect\secref{numerical_integration}, and tabulated once and for all on the grid. For cosmological problems, however, the momenta remain to be scaled according to the time dependent connection \eqnref{variable_transform}, so that $D$ needs to be re-evaluated in every time step. (A possible exception is the one of ultra-relativistic particles for which the scale invariance of $D$ can be exploited.) This shows that the entire method relies on analytic expressions for $D$. In the following section a method is presented for the computation of $D$ for arbitrary matrix elements.

\section{Expansion of the Scattering Kernel\label{sec:expansion_of_the_scattering_kernel}}

	\begin{figure}[tb]
	\begin{center}
	\footnotesize
	\input{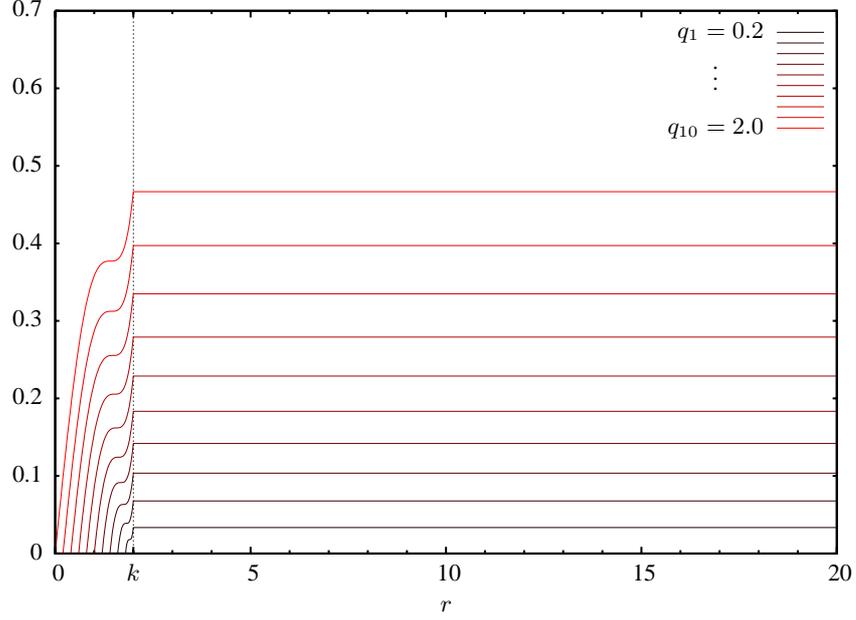}
	\end{center}
	\caption{$\Dnmind{2}{0}(2.0,p=q_i+r-2.0,q_i,r)$ for $q_i\leq k$. The flattening for $r>k$ is common to all $\Dnmind{n}{0}$. All momenta are in relative units.}
	\label{fig:Dnm_2_0qsmallerk}
	\end{figure}

	\begin{figure}[tb]
	\begin{center}
	\footnotesize
	\input{Dnm_2_0qlargerk}
	\end{center}
	\caption{This plot continues \fig\figref{Dnm_2_0qsmallerk}, $\Dnmind{2}{0}(2.0,p=q_i+r-2.0,q_i,r)$ for $q_i> k$.}
	\label{fig:Dnm_2_0qlargerk}
	\end{figure}

	\begin{figure}[tb]
	\begin{center}
	\footnotesize
	\input{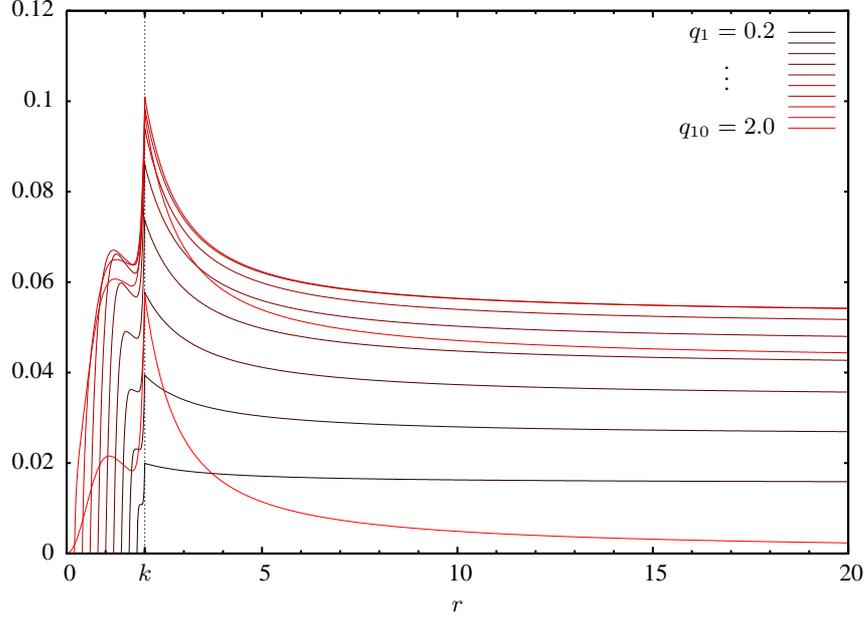}
	\end{center}
	\caption{$\Dnmind{4}{3}(2.0,p=q_i+r-2.0,q_i,r)$ for $q_i\leq k$. The kink at $r=k$ is common to all $\Dnm (k,q+r-k,q,r)$. For $r<k-q$ we have $\Dnm =0$. All momenta are in relative units.}
	\label{fig:Dnm_4_3qsmallerk}
	\end{figure}

	\begin{figure}[tb]
	\begin{center}
	\footnotesize
	\input{Dnm_4_3qlargerk}
	\end{center}
	\caption{This plot continues \fig\figref{Dnm_4_3qsmallerk}, $\Dnmind{4}{3}(2.0,p=q_i+r-2.0,q_i,r)$ for $q_i> k$.}
	\label{fig:Dnm_4_3qlargerk}
	\end{figure}

	\begin{figure}[tb]
	\begin{center}
	\footnotesize
	\input{Dnm3d}
	\end{center}
	\caption{$\Dnmind{4}{3}(2.0,p=q+r-2.0,q,r)$. \fig\figref{Dnm_4_3qsmallerk} and \fig\figref{Dnm_4_3qlargerk} correspond to cuts with $q=\const$ (the thick line corresponds to $q_9=1.8$).}
	\label{fig:Dnm3d}
	\end{figure}

The spin averaged matrix element $\meavsqu$ will in general depend on Lorentz-invariant combinations of the four-momenta of the in- and outgoing particles, usually the Mandelstam variables $s$, $t$, $u$,
\bigformularray{
s&=&(k+p)^2\kend\nonumber\\
t&=&(k-q)^2 =m_k^2+m_q^2-2E_kE_q+2\abs{\bvk}\abs{\bvq}\ctkq\kend\nonumber\\
u&=&(k-r)^2 =m_k^2+m_r^2-2E_kE_r+2\abs{\bvk}\abs{\bvr}\ctkr\dend
}{mandelstamm}

\newcommand{\betr}[1]{\left|\bvec{#1}\right|}

In the following we take $t$ and $u$ as the independent variables and $s$ is expressed by\footnote{It can be advantageous to use different variables, in which case the results of this section can be adapted easily.}:
\bigformula{s=\sum_{i=k,p,q,r}^{}m_i^2\,\,-t-u\dend}{mandelstamm_sum_relation}
In order to evaluate \eqn\eqnref{D_function} for arbitrary matrix elements we expand $\meavsqu$ in terms of $\ctkq$ and $\ctkr$:
\bigformula{\meavsqu=\sum_{n=0}^\infty\sum_{m=0}^\infty A_{nm}\left(\cos\theta_{kq}\right)^n\left(\cos\theta_{kr}\right)^m\kend}{Dnm_expansion}
Note that the coefficients $A_{nm}$ can depend on the magnitudes of the momenta. Upon integration of \eqn\eqnref{D_function} we can then write 
\bigformula{D(k,p,q,r)=\sum_{n=0}^\infty\sum_{m=0}^\infty A_{nm}(k,p,q,r)\Dnm(k,p,q,r)\kend}{Dnm_expansion_1}
assuming that the series converges for all relevant $k$, $p$, $q$ and $r$ (the momenta are still restricted by energy conservation).

In order to give a meaning to \eqn\eqnref{Dnm_expansion_1} we need to compute the integral 
\bigformularray{
\Dnm(k,p,q,r)&=& \frac{pqr}{8\pi^2}\int \dif\Omega_p\int \dif\Omega_q\int \dif\Omega_r\,\, \delta^3(\bvk+\bvp-\bvq-\bvr)(\cos\theta_{kq})^n(\cos\theta_{kr})^m\nonumber\\
&=&\frac{pqr}{64\pi^5}\int \lambda^2\dif\lambda \int e^{i\bvl\bvk}\dif\Omega_\lambda\int e^{i\bvl\bvp}\dif\Omega_p\nonumber\\
& &\hspace{10mm}\times \int e^{-i\bvl\bvq}(\cos\theta_{kq})^n\dif\Omega_q\int e^{-i\bvl\bvr}(\cos\theta_{kr})^m\dif\Omega_r\dend\nonumber\\
}{Dnm_function}

Due to this definition the $\Dnm$'s are dimensionless, scale invariant functions, i.e. \\$\Dnm(\alpha k,\alpha p,\alpha q,\alpha r)$ 
$=\Dnm(k,p,q,r)$ for any $\alpha\neq 0$. From the first line of \eqn\eqnref{Dnm_function} we can infer that, for given $k$, $p$, $q$, and $r$, the lowest order function $\Dnmind{0}{0}(k,p,q,r)$ represents an upper bound for all higher order functions $\Dnm(k,p,q,r)$.

Before investigating further the general solution, we compute $\Dnm$ for this simplest case, corresponding to $\meavsqu=1$.

We can evaluate all \dkn{solid angle integrals}, which $\meavsqu$ does not depend on, in \eqn\eqnref{Dnm_function}, using 
\bigformula{\int e^{\pm i\bvl\bvp}\dif\Omega_p=\frac{4\pi}{\lambda p}\sin(\lambda p)\dend}{solid_angle_integral}
Thus $\Dnmind{0}{0}$ simplifies to\footnote{Without loss of generality we assume $k,\,q,\,r>0$ in the following. The cases $k=0$, $q=0$, $r=0$ can be understood in the limiting sense.}
\bigformula{\Dnmind{0}{0}(k,p,q,r)=\frac{4}{k\pi}\int_0^\infty\sin(\lambda k)\sin(\lambda p)\sin(\lambda q)\sin(\lambda r)\frac{\dif\lambda}{\lambda^2}\dend}{collision_integral_reduced_constant_m}
Using the addition theorems for $\sin$ and $\cos$, the result for this integral is found to be
\bigformularray{\Dnmind{0}{0}(k,p,q,r)&=&\nonumber\\
&&\hspace{-30mm}=\frac{1}{4\,k}\,\big( \abs{k-p+q+r}- \abs{k+p-q-r} +
\abs{k+p+q-r}+\abs{k-p-q-r} \nonumber\\ &&\hspace{-21mm} +\abs{k+p-q+r}- \abs{k-p-q+r} -\abs{k-p+q-r}- \abs{k+p+q+r} \big)\kend
}{}
or equivalently
\bigformularray{
\Dnmind{0}{0}(k,p,q,r)=\frac{1}{2k}\big(\salign
\ramp{\termc{k-p+q+r}}-\ramp{\termc{k+p-q-r}}\nonumber\\
+\salign\ramp{\termc{k+p+q-r}}+\ramp{\termc{k-p-q-r}}\nonumber\\
+\salign\ramp{\termc{k+p-q+r}}-\ramp{\termc{k-p-q+r}}\nonumber\\
-\salign\ramp{\termc{k-p+q-r}}-\ramp{\termc{k+p+q+r}}
\big)\kend
}{D_0_0_result_1}
where we introduced the ramp function: 
\bigformula{\ramp{(x)}=x\,\heaviside{x}=\left\{\begin{matrix}x,\,\,\text{for}\,\,  x>0\\ 0,\,\,\text{for}\,\, x\leq 0\end{matrix}\right\}\dend}{}

Because of the proportionality of $\Dnm$ to $\termprefac{}$ from \eqn\eqnref{heaviside_prefactor}, we can infer that \\$\termc{k-p+q+r}\geq 0$, $\termc{k+p-q+r}\geq 0$, $\termc{k+p+q-r}\geq 0$ and $\termc{k-p-q-r}\leq 0$ for all values of $k$, $p$, $q$ and $r$ for which the prefactor is non-zero (none of the momenta can be greater than the sum of the other three).
Obviously, the term $\termc{k+p+q+r}$ is always positive. Introducing the abbreviations
\bigformularray{\salign c_1=k+p-q-r,\quad c_2=k-p+q-r,\quad c_3=k-p-q+r\nonumber\\ \salign \text{and}\quad R_{{1}}=\ramp{(c_1)},\quad R_{{2}}=\ramp{(c_2)},\quad R_{{3}}=\ramp{(c_3)}\kend}{ramp_and_cterm_abbreviations}
for the remaining combinations with indefinite sign, we find for \eqn\eqnref{D_0_0_result_1} the compact form:
\bigformularray{\Dnmind{0}{0}(k,p,q,r)=\termprefac{\frac{1}{2k}}\big( -R_{{1}}-R_{{2}}-R_{{3}}+2\,k \big)\dend}{Dnm_0_0_result}
From this it is obvious that $\Dnmind{0}{0}(k,p,q,r)\leq 1$ everywhere. Since the smallest value, with all the $R_i$'s being positive, is $\termprefac{}(-k+p+q+r)/(2k)\geq 0$ it is also obvious that $\Dnmind{0}{0}(k,p,q,r)\geq 0$. Remembering the note from above, we conclude that $\abs{\Dnm(k,p,q,r)}\leq\Dnmind{0}{0}(k,p,q,r)\leq 1$ for all $k$, $p$, $q$ and $r$.
This important property guarantees pointwise absolute convergence of the series \eqn\eqnref{Dnm_expansion_1} if the series of the coefficients in the expansion \eqnref{Dnm_expansion},
\formula{\sum_{n=0}^\infty\sum_{m=0}^\infty A_{nm}(k,p,q,r)\kend}{Dnm_expansion_convergence_criterion} is pointwise absolute convergent.

In case of massless particles \eqn\eqnref{Dnm_0_0_result} can be simplified further to (using energy conservation)
\bigformularray{\Dnmind{0}{0}(k,p,q,r)&=&\termprefac{\frac{1}{k}}\left(k-\ramp\left({q-k}\right)-\ramp\left({r-k}\right)\right)\nonumber\\ &=& \termprefac{\frac{1}{2k}}(q+r-\abs{q-k}-\abs{r-k})\dend}{}

We now turn to the computation of $\Dnm$ for general $n$ and $m$. In \eqn\eqnref{Dnm_function} the Fourier integrals, 
\begin{equation}
\int e^{\pm i\bvl\bvq}(\cos\theta_{kq})^n\dif\Omega_q = \int e^{\pm i\bvl\bvq}(\hat{\bvk}\hat{\bvq})^n\dif\Omega_q\kend 
\end{equation}
on the unit sphere, can be expressed as a finite series of spherical Bessel functions of the first kind (see \app\appref{bessel_functions} for the derivation):
\bigformula{\int e^{\pm i\bvl\veceta}(\hat{\bvk}\vecetaunit)^n\dif\Omega_{\hat{\veceta}}  =
4\pi (\pm i)^\indl\sum_{\indk = 0}^{\floorbr{\frac{\indl}{2}}}\coeffa{\indl}{\indk}\frac{\sbessel{\indl-\indk}{\vecetaabs\lambda}}{(2\,\vecetaabs\lambda)^\indk}(\hat{\bvk}\hat{\bvl})^{\indl -2\indk}\kend}{sbessel_expansion}
with numeric coefficients 
\bigformula{\coeffa{n}{\alpha}=\frac{(-1)^\alpha n!}{\alpha!(n-2\alpha)!} \dend}{coeffa}

Inserting \eqn\eqnref{sbessel_expansion} into \eqn\eqnref{Dnm_function} we find:
\begin{align}
\Dnm =
\frac{pqr}{\pi^2}\int & \dif\lambda \,\lambda^2 \sum_{\alpha=0}^{\floorbr{\frac{n}{2}}}\sum_{\beta=0}^{\floorbr{\frac{m}{2}}}(-i)^{n+m}\coeffa{n}{\alpha}\coeffa{m}{\beta}(2q\lambda)^{-\alpha}(2r\lambda)^{-\beta} \nonumber\\
& \times \sbessel{0}{p\lambda}\sbessel{n-\alpha}{q\lambda}\sbessel{m-\beta}{r\lambda}\int e^{i\bvl\bvk} (\cos\theta_{k\lambda})^{n+m-2(\alpha +\beta)}\dif\Omega_\lambda
\kend\nonumber\\
\label{eqn:Dnm_function_1}
\end{align}
where the inner integral is again of type \eqnref{sbessel_expansion}, such that we arrive at:
\begin{align}
\Dnm =
\frac{4pqr}{\pi} & \sum_{\alpha=0}^{\floorbr{\frac{n}{2}}}\sum_{\beta=0}^{\floorbr{\frac{m}{2}}}\sum_{l=0}^{\floorbr{\frac{n+m}{2}-(\alpha +\beta)}}(-1)^{\alpha +\beta}\coeffa{n}{\alpha}\coeffa{m}{\beta}\coeffa{n+m-2(\alpha +\beta)}{l}(2q)^{-\alpha}(2r)^{-\beta}(2k)^{-l} \nonumber\\
& \times \intIn{\alpha +\beta +l}{n+m-2(\alpha +\beta )-l}{0}{n-\alpha}{m-\beta}{k}{p}{q}{r}
\kend\nonumber\\
\label{eqn:Dnm_function_2}
\end{align}
with 
\bigformula{\intIn{n}{l_1}{l_2}{l_3}{l_4}{k}{p}{q}{r}=\int_0^\infty \lambda^{2-n} \sbessel{l_1}{k\lambda}\sbessel{l_2}{p\lambda}\sbessel{l_3}{q\lambda}\sbessel{l_4}{r\lambda}\dif\lambda\dend
}{}
Unfortunately, the remaining integral over four spherical Bessel functions is known to represent a mathematical problem itself. 
Because of the rapidly oscillating integrand it is also diffcult to access by numerical methods.

Due to Rayleigh's formula \eqnref{rayleighs_formula}, the integrand can always be decomposed into products of four sine and cosine functions and an inverse power of $\lambda$, 
\newcommand{\trig}[2]{\mbox{trig}_{#1}\left(#2\right)}
\begin{equation}
\int_0^\infty \frac{\trig{1}{k\lambda}\trig{2}{p\lambda}\trig{3}{q\lambda}\trig{4}{r\lambda}}{\lambda^n}\dif\lambda \kend\label{eqn:sin_cos_integral}
\end{equation}
where $\trig{i}{x\lambda}$ is either $\sin (x\lambda)$ or $\cos (x\lambda)$. However the integrand in \eqn\eqnref{sin_cos_integral} has a non-integrable singularity at $\lambda =0$ if the number of sines exceeds $n$.
In principle, the problem can be circumvented by performing a Laurent series expansion of the integrand in \eqn\eqnref{sin_cos_integral} and subtracting the divergent part. The finite part can then be computed for all possible combinations of the indices. Since we expect a finite overall result for \eqn\eqnref{Dnm_function_2}, the different divergent parts in the sum need to cancel.

Here we make a different approach which is based on an explicit expression for the integral 
\begin{equation}
\intIn{0}{l_1}{l_2}{l_3}{l_4}{k}{p}{q}{r}=\int_0^\infty\lambda^2\sbessel{l_1}{k\lambda}\sbessel{l_2}{p\lambda}\sbessel{l_3}{q\lambda}\sbessel{l_4}{r\lambda}\diff{\lambda}\kend \label{eqn:intI}
\end{equation}
valid, provided that there exists an integer number $L$ which satisfies the conditions:
\begin{align}
&\abs{l_1-l_2}\leq L \leq l_1+l_2\,\, \wedge \,\, \abs{l_3-l_4}\leq L \leq l_3+l_4\kend\,\,\text{and}\nonumber\\
&l_1+l_2+L \;\text{and}\; l_3+l_4+L \;\text{even}\; \label{eqn:intI_condtion}\dend
\end{align}

In order to bring the integrals $\intIn{\alpha +\beta +l}{n+m-2(\alpha +\beta )-l}{0}{n-\alpha}{m-\beta}{k}{p}{q}{r}$ into the form of \eqn\eqnref{intI} we apply the recurrence relation for spherical Bessel functions \cite{Abramowitz:1970}:
\begin{equation}
\sbessel{n}{z}=\frac{z}{2n+1}\left(\sbessel{n-1}{z}+\sbessel{n+1}{z}\right)\dend\label{eqn:bessel_recurrence}
\end{equation}
Applying this relation $r$ times with respect to $\sbessel{n}{x\lambda}$ yields a sequence of spherical Bessel functions of order $n-r,\,n-r+2 \ldots n+r-2,\,n+r$ and an overall prefactor $(x\lambda)^r$. Therefore, we apply \eqn\eqnref{bessel_recurrence} $l$ times with respect to $\sbessel{n+m-2(\alpha +\beta )-l}{k\lambda}$, $\alpha$ times with respect to $\sbessel{n-\alpha}{q\lambda}$ and $\beta$ times with respect to $\sbessel{m-\beta}{r\lambda}$ in \eqn\eqnref{Dnm_function_1}. This leads to a set of integrals of type $\intIn{0}{l_1}{l_2}{l_3}{l_4}{k}{p}{q}{r}$, where $0\leq n+m-2(\alpha +\beta+l)\leq l_1\leq n+m-2(\alpha +\beta)$, $l_2 =0$, $0\leq n-2\alpha\leq l_3\leq n$ and $0\leq m-2\beta\leq l_4\leq m$.
These integrals can be evaluated according to \cite{Mehrem:1991} if the conditions \eqnref{intI_condtion} on the indices are met. 

Since different authors found different, relevant expressions for integrals involving three Bessel functions \cite{jackson:1972,Anni:1974,Elbaz:1974}, we repeat here the derivation for integrals involving four spherical Bessel functions from the former ones.

With help of the closure relation for spherical Bessel functions \eqnref{spherical_bessel_closure_relation}, integrals of the form \eqnref{intI} can be reduced to integrals of three spherical Bessel functions:
\bigformularray{
\intIn{0}{l_1}{l_2}{l_3}{l_4}{k}{p}{q}{r} &=& \frac{2}{\pi}\int_0^\infty\dif\lambda\, \lambda^2\int_0^\infty z^2j_{l_1}(kz)j_{L}(\lambda z)j_{l_2}(pz)\dif z\nonumber\\ & &
\times \int_0^\infty z'^2j_{l_1}(qz')j_{L}(\lambda z')j_{l_2}(rz')\dif z'\nonumber\\ & = &
\frac{2}{\pi}\int_0^\infty \dif\lambda\,\lambda^2\intIthree{l_1}{L}{l_2}{k}{\lambda}{p}\intIthree{l_3}{L}{l_4}{q}{\lambda}{r}\kend
}{int_four_bessel_to_int_three_bessel}
defining
\bigformula{\intIthree{l_1}{l_2}{l_3}{k}{p}{q}=\int_0^\infty\,\lambda^2 \sbessel{l_1}{k\lambda}\sbessel{l_2}{p\lambda}\sbessel{l_3}{q\lambda}\dif\lambda\dend}{}
Inserting the expression \eqnref{intIthree_result} for $\intIthree{l_1}{l_2}{l_3}{k}{p}{q}$ found by Mehrem et al. \cite{Mehrem:1991} and performing the integration (after inserting the explicit representation \eqnref{legendre_poly_explicit} for the Legendre polynomials) yields
\bigformularray{
& & \hspace{-10mm}\intIn{0}{l_1}{l_2}{l_3}{l_4}{k}{p}{q}{r} = \nonumber\\ & &
(-1)^L\frac{\pi i^{l_1-l_2+l_3-l_4}}{8kpqr}\sqrt{(2l_2+1)(2l_4+1)}\,\termprefac{}\bigg(\frac{k}{p}\bigg)^{l_2}\bigg(\frac{q}{r}\bigg)^{l_4}\nonumber\\ & &  \times \wignerthreej{l_1}{l_2}{L}{0}{0}{0}^{-1}\wignerthreej{l_3}{l_4}{L}{0}{0}{0}^{-1}\sum_{n=0}^{l_2}\sum_{n'=0}^{l_4}\left(\begin{pmatrix}2l_2\nonumber\\ 2n\end{pmatrix}\begin{pmatrix}2l_4\nonumber\\ 2n'\end{pmatrix}\right)^{\frac{1}{2}}
\nonumber\\ & & \times \sum_{l=\abs{l_1 - l_2 + n}}^{l_1 + l_2 - n}\sum_{l'=\abs{l_3 - l_4 + n'}}^{l_3 + l_4 - n'}(2l+1)(2l'+1)\wignerthreej{l_1}{l_2-n}{l}{0}{0}{0}\wignerthreej{l_3}{l_4-n'}{l'}{0}{0}{0}\nonumber\\ & & \times\wignerthreej{L}{n}{l}{0}{0}{0}\wignerthreej{L}{n'}{l'}{0}{0}{0}\wignersixj{l_1}{l_2}{L}{n}{l}{l_2-n}\wignersixj{l_3}{l_4}{L}{n'}{l'}{l_4-n'}\nonumber\\ &&\times\frac{J(k,p,q,r;n,n',l,l')}{k^nq^{n'}} \kend
}{intfour_result}
where 
\bigformula{\wignerthreej{j_1}{j_2}{j_3}{m_1}{m_2}{m_3}\,\,\text{and}\,\,\wignersixj{j_1}{j_2}{j_3}{j_4}{j_5}{j_6}}{} denote the Wigner 3j- and 6j-symbols, respectively (these are purely numeric factors related to the Clebsch-Gordan coefficients and Racah's W-coefficients, respectively \cite{Edmonds:1954}) and with
\bigformularray{
&&\hspace{-10mm} J(k,p,q,r;n,n',l,l')  = \nonumber\\ &  &=\sum_{s=0}^{\floorbr{\frac{l}{2}}}\sum_{t=0}^{\floorbr{\frac{l'}{2}}}A_{l,s}A_{l',t}(2k)^{2s-l}(2q)^{2t-l'} \sum_{\mu=0}^{l-2s}\sum_{\nu=0}^{l'-2t}\left(\begin{matrix}l-2s \mu\end{matrix}\right)\left(\begin{matrix}l'-2t  \nu\end{matrix}\right)\nonumber\\ &&\quad\quad\times (k^2-p^2)^{l-2s-\mu}(q^2-r^2)^{l'-2t-\nu}\funcU{n+n'+2\mu+2\nu+2s+2t-l-l'+1}{k}{p}{q}{r}\kend\nonumber\\
}{intfour_result_with}

where we have defined \bigformula{\funcU{\alpha}{k}{p}{q}{r}=\frac{\Min{k+p,q+r}^{\alpha}-\Max{\abs{p-k},\abs{r-q}}^{\alpha}}{\alpha}\dend}{funcU}
The expected prefactor $\termprefac{}$ in \eqn\eqnref{intfour_result} stems from the integration of the Heaviside step function in \eqn\eqnref{intIthree_result}.

From \eqn\eqnref{intIthree_result} the result \eqnref{intfour_result} also inherits the restrictions on the indices of the Bessel functions $\abs{l_1-l_2}\leq L \leq l_1+l_2 \wedge \abs{l_3-l_4}\leq L \leq l_3+l_4$. Note, that, in general, \eqn\eqnref{intfour_result} can be evaluated for different values of $L$ and with different mapping of the indices $l_1$, $l_2$, $l_3$ and $l_4$, leading to different equivalent results.

All sums in \eqn\eqnref{Dnm_function_2}, \eqn\eqnref{intfour_result} and \eqn\eqnref{intfour_result_with} are finite, so they can be used to determine the functions $\Dnm$. In order to demonstrate the usefulness of this result we apply it explicitly to the cases of $\Dnmind{0}{0}$ and $\Dnmind{2}{0}$.
Evaluating \eqnref{Dnm_function_2} for $n=m=0$ we find
\bigformula{\Dnmind{0}{0}={\frac {4\,p q r}{\pi }}\intIn{0}{0}{0}{0}{0}{k}{p}{q}{r}\dend
}{}
Applying \eqn\eqnref{intfour_result} leads immediately to
\bigformularray{
\Dnmind{0}{0}&=&{\frac {\termprefac{} \funcU{1}{k}{q}{p}{r} }
{2\,k}}\nonumber\\ &=& {\frac {\termprefac{} \left( \min \left( p+r,k+q \right) -\max
 \left(  \left| -q+k \right| , \left| -r+p \right|  \right)  \right) }
{2\,k}}\dend}{D00_result}
Differentiating the eight cases with $\sgn{(c_i)}=\pm 1$, this can be shown to be equivalent to \eqn\eqnref{Dnm_0_0_result}.

A more sophisticated example is $\Dnmind{2}{0}$. Again, from \eqn\eqnref{Dnm_function_2}, we find
\bigformularray{
\Dnmind{2}{0}&=&{\frac {8\,p q r}{\pi }}\bigg(
\frac{1}{2}\,\intIn{0}{2}{0}{2}{0}{k}{p}{q}{r}-\nonumber\\&&\quad\quad\quad\quad-{\frac {
\intIn{-1}{1}{0}{2}{0}{k}{p}{q}{r}}{2\,k}}+{\frac {
\intIn{-1}{0}{0}{1}{0}{k}{p}{q}{r}}{2\,q}}\bigg)
\dend}{}
We apply the recurrence relation \eqnref{bessel_recurrence} To the second and third term with respect to $\sbessel{1}{k\lambda}$ and $\sbessel{1}{q\lambda}$ respectively. This leads to 
\bigformula{\Dnmind{2}{0}={\frac {8\,p q r}{\pi }}\bigg(\frac{1}{6}\,\intIn{0}{0}{0}{0}{0}{k}{p}{q}{r}+\frac{1}{3}\,
\intIn{0}{2}{0}{2}{0}{k}{p}{q}{r}
\bigg)\dend
}{}
The terms involving $\intIn{0}{0}{0}{2}{0}{k}{p}{q}{r}$ did cancel exactly and both of the remaining integrals can be evaluated according to \eqn\eqnref{intfour_result} (with the unique choice of $L=0$ and $L=2$ for the first and the second integral, respectively), giving after some algebra:
\bigformularray{\Dnmind{2}{0}&=&\frac{\termprefac{}}{8\,{q}^{2}{k}^{3}}\nonumber\\&&\times{ {\left( \left( {k}^{2}+{q}^{2}
 \right) ^{2}\funcU{1}{k}{q}{p}{r}-2\,\left( {k}^{2}+{q}^{2} \right) 
\funcU{3}{k}{q}{p}{r} +\funcU{5}{k}{q}{p}{r}\right) }}\dend\nonumber\\}{D20_result}

Eqn.\,\eqnref{D20_result} as well as all other functions $\Dnm (k,p,q,r)$ can be brought into the compact form 
\newcommand{\termpolypref}[1]{B_{#1}}
\newcommand{\termremain}{C}
\newcommand{\termoverall}{A}
\bigformularray{\Dnmind{n}{m}(k,p,q,r)=\termoverall\frac{\termprefac{}}{k^{n+m+1}q^n r^m}\big( \termpolypref{1}R_{{1}}+\termpolypref{2}R_{{2}}+\termpolypref{3}R_{{3}}+C\big)\dend}{Dnm_compact_notation}
We computed the numeric prefactor $\termoverall$ and the momentum dependent coefficients, $\termpolypref{1}(k,p,q,r)$, $\termpolypref{2}(k,p,q,r)$, $\termpolypref{3}(k,p,q,r)$ and $\termremain(k,p,q,r)$ for $n+m\leq 16$. They are listed in \app\appref{functions_Dnm} for $\Dnm$ with $n+m\leq 5$.\footnote{They become lengthy for greater indices.} The coefficients $\termpolypref{i}$ and $\termremain$ are homogeneous multivariate polynomials of degree $2\,(n+m)$ and $2\,(n+m)+1$ in $k$, $p$, $q$, $r$. The number of elementary operations, necessary to evaluate $\Dnm$, increases with increasing $n$ and $m$, however their shape permits considerable optimisation, especially when many $\Dnm$'s are to be computed. In addition, when dealing with networks of Boltzmann equations, they can be used for all matrix elements in the system. 

Fig.\,\figref{Dnm_2_0qsmallerk} and \fig\figref{Dnm_2_0qlargerk} show $\Dnmind{2}{0}(2.0,p=q_i+r-2.0,q_i,r)$ plotted against $r$ (all momenta are in relative units) for various values of $q_i$. For $r>k$ the graph of $\Dnmind{2}{0}$ becomes constant. This can be understood by the observation that $\funcU{\alpha}{k}{q+r-k}{q}{r}$ is independent of $r$ for $r>k$. The same relation holds for all other $\Dnmind{n}{0}$'s (and a corresponding one for $\Dnmind{0}{m}$). Fig.\,\figref{Dnm_4_3qsmallerk} and \fig\figref{Dnm_4_3qlargerk} show similar plots for $\Dnmind{4}{3}$ which do not have this property. Fig.\,\figref{Dnm3d} shows an surface plot of $\Dnmind{4}{3}$ for fixed $k$, as a function of $q$ and $r$, $\Dnmind{4}{3}(2.0,p=q+r-2.0,q,r)$.

In general the shape of the graphs varies strongly with varying indices $n$ and $m$. All functions possess a kink at $r=k$, because \bigformularray{\Dnmind{n}{m}(k,p=q+r-k,q,r)=\termoverall\frac{\termprefac{}}{k^{n+m+1}q^n r^m}\big( 2\,\termpolypref{2}\ramp{(k-r)}+2\,\termpolypref{3}\ramp{(k-q)}+C\big)\kend}{Dnm_compact_notation_1} and the properties $\Dnm(k,p,q,r) =0$ for $r<k-q$, ($\termprefac{}\overset{p=q+r-k}{=}\heaviside{q+r-k}=\heaviside{p}$) and $\underset{r\rightarrow 0+}{\lim} \Dnm(k,p,q,r) =0$ for $k$, $p$, and $q$ held constant. The later one can be inferred from $\abs{\Dnm(k,p,q,r)}\leq\Dnmind{0}{0}(k,p,q,r)$ and $\underset{r\rightarrow 0+}{\lim} \Dnmind{0}{0}(k,p,q,r) =0$, which is obvious from \eqn\eqnref{D00_result}. Similar relations hold for the $q$ dependence (with $k$ and $r$ fixed) and in the case of massive particles.

\section{Numerical Integration\label{sec:numerical_integration} of \texorpdfstring{\mathversion{bold}${D}$\mathversion{normal}}{D}}
In this section we derive a formula for $D(k,p,q,r)$, suitable for numerical integration of arbitrary matrix elements. Again, we  assume that the angular dependence of $\meavsqu$ is given in terms of $\ctkq$ and $\ctkr$, i.e. in terms of the momentum transfer $t$ and $u$. A possible dependence on $s$ can be expressed in terms of $t$ and $u$ exploiting energy and momentum conservation, \eqn\eqnref{mandelstamm_sum_relation}. 

We orientate the coordinate system such that the $z$-axis points in the direction of $\bvk$. We can then write
\bigformularray{
\cttwovec{\lambda}{v}&=&\hat{\bvec{\lambda}}\cdot\hat{\bvec{v}}=\ctonevec{\lambda}\ctonevec{v}+\stonevec{\lambda}\stonevec{v}\cos(\phi_\lambda-\phi_v)\kend\nonumber \\
 \ctkq&=&\ctq\kend \nonumber \\
 \ctkr&=&\ctr\dend}{cosine_relations}
Using \eqn\eqnref{sbessel_expansion} for the $\Omega_p$ integration we can write \eqn\eqnref{D_function} as
\bigformula{
D(k,p,q,r)=\frac{pqr}{16\pi^4}\int \dif^3\lambda\,\,e^{i\bvl\bvk}\frac{\sin\left(\lambda p\right)}{\lambda p}\int e^{-i\bvl\bvq}\dif\Omega_q\int e^{-i\bvl\bvr}\dif\Omega_r\meavsqu\dend
}{D_function_integrated_num_1}

We can then perform the integration over $\phi_q$ and $\phi_r$ since $\meavsqu$ does not depend on these angles
\bigformularray{
\int e^{-i\bvl\bvr}\dif\Omega_r\meavsqu =\nonumber\\ &&\hspace{-30mm} =\int_{-1}^{1} e^{-i\lambda r \ctl\ctr}\dif\ctr\int_{0}^{2\pi} e^{-i\lambda r \stl\str\cos\left(\phi_\lambda -\phi_r\right)}\meavsqu \dif{\phi_r}\nonumber \\
&&\hspace{-30mm}= \int_{-1}^{1} e^{-i\lambda r \ctl\ctr}\meavsqu \dif\ctr \int_{0}^{\pi} 2\cos\left(\lambda r \stl\str\cos\left(\phi_r\right)\right) \dif{\phi_r}\nonumber \\
&&\hspace{-30mm} = 2\pi \int_{-1}^{1} e^{-i\lambda r \ctl\ctr} J_0(\lambda r\stl\str)\meavsqu\dif\ctr\kend }{plane_wave_bessel}
where we have taken into account the fact that the inner integral does not depend on $\phi_\lambda$, since the integration is over $2\pi$ and the odd symmetry of the imaginary part of this integral with respect to $\phi_r$. The remaining integral was recognized as the integral definition of $J_0$, the Bessel function of the first kind of order zero, see e.g. \cite{Abramowitz:1970}.

Inserting \eqn\eqnref{plane_wave_bessel} into \eqn\eqnref{D_function_integrated_num_1} gives
\bigformula{D(k,p,q,r)=\frac{pqr}{4\pi^2}\int_0^\infty \lambda^2\frac{\sin{\left(\lambda p\right)}}{\lambda p} I \dif\lambda\kend}{D_function_integrated_num_2}
with
\begin{eqnarray}
I&=&\int e^{i\lambda k \ctl}\dif\Omega_\lambda\int_{-1}^{1} e^{-i\lambda q \ctl\ctq}\dif\ctq\int_{-1}^{1} e^{-i\lambda r \ctl\ctr}\dif\ctr\nonumber \\
& &\times J_0(\lambda q\stl\stq) J_0(\lambda r\stl\str) \meavsqu\dend\nonumber \label{eqn:integral_J}\\
\end{eqnarray}
For the product of the two Bessel functions we may use the relation \eqnref{formula_prod_two_bessel}.
Inserting it into \eqn\eqnref{integral_J} and interchanging the order of integration, we find
\bigformula{I=2\pi\int_{-1}^{1}\int_{-1}^{1}\dif \ctq\dif \ctr \meavsqu I'\kend}{integral_I_1}
with
\bigformularray{I' &= & \frac{1}{\pi} \int_0^\pi\int_0^\pi e^{i\lambda\left(k-q\ctq-r\ctr\right)\ctl} \nonumber\\
& &\times J_0\left(\lambda\stl\sqrt{\left(q\stq\right)^2+\left(r\str\right)^2-2qr\stq\str\cos\left(x\right)}\right)\stl\dif x\dif\theta_\lambda \dend\nonumber\\}{integral_I_prime}
Again interchanging the order of the integrals and exploiting the odd symmetry of the imaginary part of the exponential, we get
\bigformularray{I'& = & \frac{1}{\pi} \int_0^\pi\int_0^\pi \cos\left({\lambda\left(k-q\ctq-r\ctr\right)\ctl}\right) \nonumber\\
& &\times J_0\left(\lambda\stl\sqrt{\left(q\stq\right)^2+\left(r\str\right)^2-2qr\stq\str\cos\left(x\right)}\right)\stl\dif\theta_\lambda\dif x\kend \nonumber\\ }{integral_I_prime_1}
Now we apply the integral \eqnref{formula_cos_bessel_sin} to arrive at
\bigformula{I'=\frac{2}{\pi}\int_0^\pi\frac{\sin\left(\lambda \sqrt{f(x)}\right)}{\lambda \sqrt{f(x)}}\dif x\kend}{integral_I_prime_2}
with
\bigformula{f(\cos x)=\left(k-q\ctq-r\ctr\right)^2+\left(q\stq\right)^2+\left(r\str\right)^2-2\,qr\stq\str\cos x \kend}{definition_f}
considering $0\leq\theta_q,\,\,\theta_r\leq \pi$ as parameters.
If we interpret $x$ as the polar angle enclosed by $\bvq$ and $\bvr$ we can write,
\bigformularray{f(\cos x) &=& (\bvk-\bvq )^2+(\bvk -\bvr )^2-(\bvq -\bvr )^2-k^2+q^2+r^2\nonumber\\ &=& (\bvk-\bvq )^2+(\bvk -\bvr )^2-(\bvq -\bvr )^2-k^2+q^2+r^2 + (E_k+E_p-E_q-E_r)^2\nonumber\\ &=& s+t+u-\sum_i m_i^2 +p^2\kend}{f_relation}
under the assumption of energy and momentum conservation.

Since $qr\stq\str\geq 0$ for all $\theta_q,\,\,\theta_r$, the function $f(x)$ takes its minimum for $\cos x=1$:
\bigformula{f(1)=\left(k-q\ctq-r\ctr\right)^2+\left(q\stq - r\str\right)^2\geq 0\dend}{function_f_always_pos}
$I'$ is therefore well defined.

Inserting both, \eqnref{integral_I_1} and \eqnref{integral_I_prime_2}, into \eqn\eqnref{D_function_integrated_num_2} and again exchanging the order of integration, we find:
\bigformularray{D(k,p,q,r)&=&\frac{pqr}{\pi^2}\int_{-1}^{1}\int_{-1}^{1}\dif \ctq\dif \ctr \nonumber\\ &&\quad\quad\quad\times\quad\meavsqu\int_0^\pi\dif x\int_0^\infty \lambda^2\frac{\sin{\left(\lambda p\right)}}{\lambda p} \frac{\sin\left(\lambda \sqrt{f(\cos x)}\right)}{\lambda \sqrt{f(\cos x)}} \dif\lambda\dend}{D_function_integrated_num_3}
In the rightmost integral we recognize the closure relation for spherical Bessel functions \eqnref{spherical_bessel_closure_relation} for $n=0$.

This leads to 
\bigformula{D(k,p,q,r)=\frac{qr}{2\pi\, p}\int_{-1}^{1}\int_{-1}^{1}\dif \ctq\dif \ctr \meavsqu I'' \kend}{D_function_integrated_num_4}
where we defined
\bigformula{I''=\int_0^\pi\delta\left( p-\sqrt{f(\cos x)}\right)\dif x=\int_{-1}^{1}\frac{\delta\left( p-\sqrt{f({y})}\right)}{\sqrt{1-y^2}}\dif y=\frac{2\,p\,\heaviside{F(\theta_q ,\theta_r)}}{\sqrt{F(\theta_q ,\theta_r)}}\kend}{integral_I_prime_prime}
with 
\bigformularray{F(\theta_q ,\theta_r)&=&\left(p^2-f(1)\right)\left(f(-1)-p^2\right)\nonumber\\
&&\hspace{-15mm}=\left( 2qr\stq\str\right)^2-\left[\left(k-q\ctq-r\ctr\right)^2+\left(q\stq\right)^2+\left(r\str\right)^2-p^2\right]^2\dend\nonumber\\}{definition_F}
Because of \eqn\eqnref{f_relation} the $\delta$-function in \eqn\eqnref{integral_I_prime_prime} ensures energy and momentum conservation.

The final expression for $D$ reads:
\bigformula{D(k,p,q,r)=\frac{qr}{\pi} \int_A \frac{\meavsqu}{\sqrt{F(\theta_q ,\theta_r)}} \dif \ctq\dif \ctr \kend}{D_function_integrated_num_final}
where the domain of integration $A$ is given by $-1\leq\cos\theta_q,\,\cos\theta_r\leq 1$ and $F(\theta_q ,\theta_r)> 0$.
Note, that the term $1/\sqrt{F(\theta_q ,\theta_r)}$ seems to be absent in an analogous expression in \cite{Hannestad:1995rs}. We are confident that it must be there, because in the simple case of $\meavsqu =1$ the result would be $4\,qr/\pi$, without $F$, which is different from \eqn\eqnref{Dnm_0_0_result}.

The expression \eqnref{D_function_integrated_num_final} for $D$ has only a two-dimensional integral and is by far superior to \eqn\eqnref{D_function_integrated_num_1} with respect to numerical integration. The integrand may have singular points at the boundary of $A$. Hence the routines for numerical integration must be chosen adequately. Usually, for a numerical method, it is sufficient to know $D(k,p=\sqrt{(E_q+E_r-E_k)^2-m_p^2},q,r)$ for a finite set of momenta $\{k_i,q_j,r_l\}$ on a grid. Therefore it is possible, in principle, to tabulate $D$ through numerical integration of \eqn\eqnref{D_function_integrated_num_final}. As pointed out above, for applications in cosmology, this relation is only of restricted use, since the momenta or, equivalently, the particle masses are scaled in each step of the time evolution, so that the values of $D(k_i,p,q_j,r_l)$ need to be recomputed permanently.

\section{Convergence of the Method}

	\begin{figure}[tb]
	\begin{center}
	\footnotesize
	\input{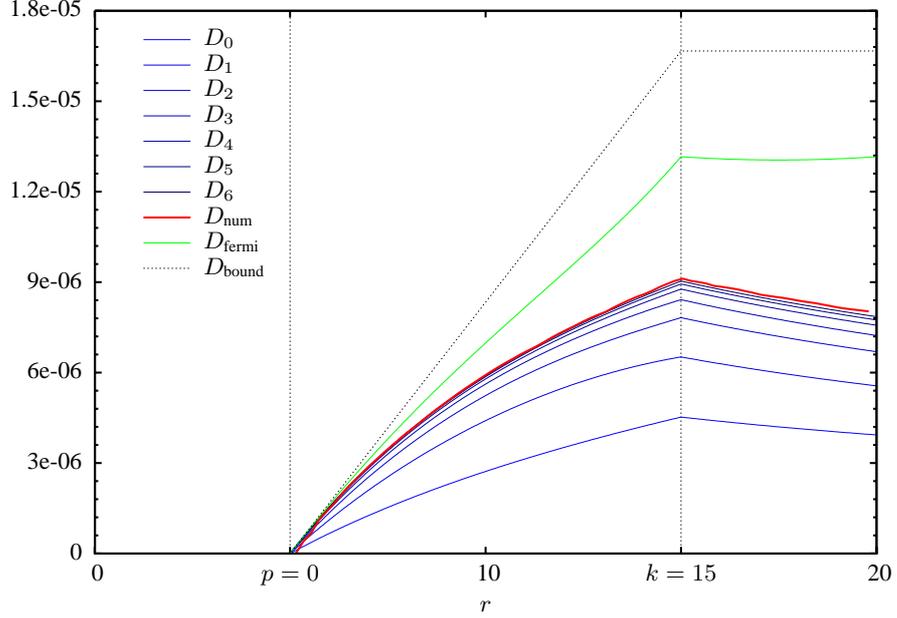}
	\end{center}
	\caption{$D(k,q+r-k,q,r)$ with $m_X=20.0$, $k=15.0$, $q=10.0$. Exact numerical result $D_{\text{num}}$, result in large $m_X$ limit $D_{\text{fermi}}$, successive analytic approximations $D_0\ldots D_6$ ($D_{n+m}$ corresponds to an expansion of $\meavsqu$ up to order $n+m$ in the cosines).}
	\label{fig:tuchannel_approx_k15q10m20}
	\end{figure}

	\begin{figure}[tb]
	\begin{center}
	\footnotesize
	\input{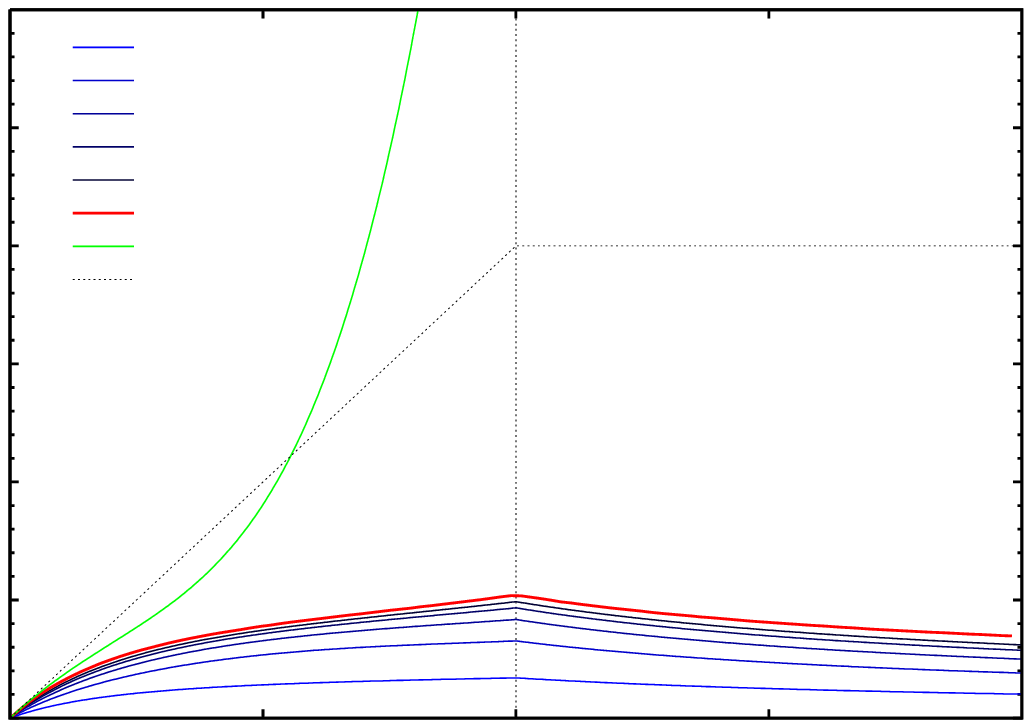}
	\end{center}
	\caption{$D(k,q+r-k,q,r)$ with $m_X=10.0$, $k=10.0$, $q=15.0$. Exact numerical result $D_{\text{num}}$, result in large $m_X$ limit $D_{\text{fermi}}$, successive analytic approximations $D_0,D_2\ldots D_8$.}
	\label{fig:tuchannel_approx_k10q15m10}
	\end{figure}

To demonstrate the application and convergence of the method described above, we apply it to a (hypothetical) matrix element involving tree-level $t$- and $u$-channel contributions. We compare the exact numerical result (by the formula derived in \sect\secref{numerical_integration}) with the approximate analytical result according to the truncated series expansion of \sect\secref{expansion_of_the_scattering_kernel}.

Without specifying a theory we take the matrix element to be (for simplicity we assume that all in- and outgoing particles are massless):
\bigformula{\meavsqu= \left( \frac{g}{{m_X}^2-t}+\frac {g}{{m_X}^2-u} \right)^{2} = \left( {\frac {g}{2\,kq}\frac{1}{ \left( a-\ctq \right) }}+{\frac{g}{2\,kr}\frac{1}{
 \left( b-\ctr \right) }} \right) ^{2}\kend
}{matrix_element_example}
with $a=1+{{m_X}^2}/{(2\,kq)}>1$ and $b=1+{{m_X}^2}/{(2\,kr)}>1$ and some mass $m_X$ of the intermediate state and a coupling $g$ with mass dimension $2$.

\newcommand{\varx}{\ctq}
\newcommand{\vary}{\ctr}
A Taylor series expansion in $\ctq$ and $\ctr$ leads to:
\bigformularray{\meavsqu &=&\frac{g^2}{4\,{k}^{2}{q}^{2}}\,\frac{1}{a^{2}}\, \left( 1+{\frac {2\,\varx}{{a}}} + \ldots\right)  + \frac{g^2}{4\,{k}^{2}{r}^{2}}\,\frac{1}{b^{2}}\, \left( 1+{\frac {2\,\vary}{{b}}} + \ldots\right) + \nonumber\\ &&+\frac{g^2}{2\, {k}^{2}{q}{r}}\,{\frac {1}{ab}}\, \left( 1+{\frac {\varx}{{a}}}+{\frac {\vary}{{b}
}}+\frac{(\varx)^2}{a^2}+\frac{\vary\varx}{ab}+\frac{(\vary)^2}{b^2 }+\ldots \right)\dend\nonumber\\
}{matrix_element_example_1}
The angle-integrated matrix element $D(k,p,q,r)$ is found by substituting every appearance of $(\ctq)^n(\ctr)^m$ by $\Dnm$:
\bigformularray{\meavsqu &=&\frac{g^2}{4\,{k}^{2}{q}^{2}}\,\frac{1}{a^{2}}\, \left( 1+{\frac {2\,\Dnmind{1}{0}}{{a}}} + \ldots\right)  + \frac{g^2}{4\,{k}^{2}{r}^{2}}\,\frac{1}{b^{2}}\, \left( 1+{\frac {2\,\Dnmind{0}{1}}{{b}}} + \ldots\right) + \nonumber\\ &&+\frac{g^2}{2\, {k}^{2}{q}{r}}\,{\frac {1}{ab}}\, \left( 1+{\frac {\Dnmind{1}{0}}{{a}}}+{\frac {\Dnmind{0}{1}}{{b}
}}+\frac{\Dnmind{2}{0}}{a^2}+\frac{\Dnmind{1}{1}}{ab}+\frac{\Dnmind{0}{2}}{b^2 }+\ldots \right)\dend
}{matrix_element_example_1a}
 Setting $\ctq =1$ and $\ctr =1$ yields the series of coefficients which converges to
\bigformula{\left.\meavsqu\right|_{\ctq =\ctr =1} = \left( {\frac {g}{2\,kq}\frac{1}{ \left( a-1 \right) }}+{\frac{g}{2\,kr}\frac{1}{
 \left( b-1 \right) }} \right) ^{2}\dend
}{matrix_element_example_2}
According to what has been said above $\left.\meavsqu\right|_{\ctq =\ctr =1}\Dnmind{0}{0}(k,p,q,r)$ represents an upper bound to $D(k,p,q,r)$. On the other hand, in the low energy limit, \eqn\eqnref{matrix_element_example} can be expanded in terms of inverse powers of ${m_X}^2$: 
\bigformula{\meavsqu\simeq\frac{g^2}{m_X^4}\left(4+4\,{\frac {t}{{{\it m_X}}^{2}}}+4\,{\frac {u}{{{\it m_X}}^{2}}}+5\,{
\frac {{t}^{2}}{{{\it m_X}}^{4}}}+2\,{\frac {tu}{{{\it m_X}}^{4}}}+5\,{
\frac {{u}^{2}}{{{\it m_X}}^{4}}}\right)\kend
}{matrix_element_example_3}
which corresponds to the Fermi-approximation and includes only $(\ctq)^n(\ctr)^m$ terms with $n+m\leq 2$ at this order (i.e. these terms can be integrated as in the previous literature).

\fig\figref{tuchannel_approx_k15q10m20} and \fig\figref{tuchannel_approx_k10q15m10} show the graphs of the exact numerical result (\eqn\eqnref{D_function_integrated_num_final}), the graphs for the ``Fermi-approximation'' (\eqn\eqnref{matrix_element_example_3}), the graphs of the theoretical upper bound (\eqn\eqnref{matrix_element_example_2}) and the graphs corresponding to successive approximations according to \eqn\eqnref{matrix_element_example_1a} for two sets of parameters. Fig.\,\figref{tuchannel_approx_k15q10m20} and \fig\figref{tuchannel_approx_k10q15m10} show that for parameters, for which the expansion in inverse powers of ${m_X}^2$ naturally fails, the approximation by the truncated $\Dnm$ expansion gives quite good results.

The rate of convergence of successive $D_i$'s torwards the exact result depends on the momenta since the coefficients in the expansion \eqnref{matrix_element_example_1} are momentum dependent. In this case it becomes worse for ${m_X}^2/(2\,kq)\ll 1$ and ${m_X}^2/(2\,kr)\ll 1$.

\section{Conclusion\label{sec:conclusion}}

In state-of-the-art computations in astroparticle physics usually all species apart from the neutrinos, which experience only weak interactions, are assumed to be in exact kinetic equilibrium and heterogeneous (at best) networks of Boltzmann- and rate equations are solved instead of the full system of kinetic equations. The number of species involved in realistic systems is large and requires a unified treatment of the different particles and interactions.

A method for the solution of the space homogeneous Boltzmann equation (with isotropic distribution function), for general scattering laws was presented here. The method relies on the expansion of the matrix element in terms of cosines of two ``scattering angles''. For the separate terms in this expansion the full angular integration was carried out. The functions $\Dnmind{0}{0}, \Dnmind{0}{1}$ and $\Dnmind{0}{2}$, corresponding to matrix elements in Fermi-Approximation, were used in previous literature to compute non-equilibrium corrections to the neutrino-distribution functions and have been obtained in the lowest order of the expansion.

Though our starting point was the relativistic form of the Boltzmann equation, as encountered in astroparticle physics, the method can be used for the non-relativistic equation as well. In any case, it allows for the full angular integration of the scattering kernel, reducing the collision integral from effective dimension $5$ to dimension $2$. The only prerequisite is that the matrix element can be expanded into a series of the scattering angles and that this series converges rapidly enough. The quantum statistical terms for blocking and stimulated emission can be carried along.

In the introduction we mentioned that, at high densities and temperatures, modifications of the Boltzmann equation might become necessary (if these are sufficient at all). One such modification, which has been suggested on various occasions, is the inclusion of higher order scattering processes.
Since the representation of the angular integral in terms of spherical Bessel functions (\ref{eqn:Dnm_function_1}-\ref{eqn:Dnm_function_2}) and the integral \eqnref{intfour_result}, by means of \eqn\eqnref{int_four_bessel_to_int_three_bessel}, can be generalized for higher order processes (in which case more than four Bessel functions appear in the integrals) the method, described above, can in principle be used to reduce the corresponding collision integrals from dimension $3\,(n-1)$ to $n-2$. ($n$ is the number of particles involved)

The functions $\Dnm$, though of very simple structure, can become lengthy for higher orders. Moreover, due to the presence of very different relaxation time-scales the system of ODE's, corresponding to the numerical method  presented in \sect\secref{a_simple_numerical_model}, tends to behave stiff. Therefore, in possible implementations, careful optimisation for efficiency and stability is necessary. 

Let us add, that the expansion of the scattering kernel in terms of the cosines of the angles is not a new idea. For example in \cite{Kuegerl:1989} an expansion of the scattering kernel has been combined with a moment method for the non-relativistic, inhomogeneous Boltzmann equation. The expansion of generic kernels with full integration of the angular part, in the space-homogeneous and isotropic case, seems to be new, however.

\subsection*{Acknowledgements}
\noindent Andreas Hohenegger was supported by the ``Sonderforschungsbereich'' TR27.

\newpage

\begin{appendix}

\renewcommand{\theequation}{A.\arabic{equation}}

\setcounter{equation}{0}

\section{Reduction of \texorpdfstring{\mathversion{bold}$\mathbf{C^{1\leftrightarrow 2}}$\mathversion{normal}}{C1-2} like Collision Integrals\label{app:C12_like_collision_integrals}}
For collision integrals describing decays and inverse decays the angle integrated matrix element can be defined similarly to \eqn\eqnref{Dnm_function}. In this case the matrix element is a constant and only the zeroth-order integral, corresponding to $\meavsqu =1$, has to be computed. The collision integral reads
\bigformularray{
C^{1\leftrightarrow 2}\left[f\right]&=&\frac{1}{2E_k}\int (2\pi)^4\delta^{(4)}(k-q-r)\meavsqu F'[f]\,\frac{\dif^3q}{(2\pi)^32E_q}\frac{\dif^3r}{(2\pi)^32E_r}\kend
}{collision_integral_decays_1}
with $F'[f]=(1-\xi f_k)f_qf_r-f_k(1-\xi f_q)(1-\xi f_r)$.
Performing the same steps as in the main text, we derive
\bigformula{C^{1\leftrightarrow 2}\left[f\right]=\frac{1}{32\pi E_k}\int\heaviside{E_q-m_q} F'[f] D'(k,q,r)\frac{r\dif r}{E_r}\kend}{boltzmann_equation_decay_reduced} 
where $E_q=E_k-E_r$, $q=\sqrt{E_q^2-m_q^2}$ and we have defined the function $D'$ as 
\bigformula{D'(k,q,r)=\frac{qr}{8\pi^4}\int \lambda^2\dif\lambda\int e^{i\bvl\bvk}\dif\Omega_\lambda\int e^{-i\bvl\bvq}\dif\Omega_q\int e^{-i\bvl\bvr}\dif\Omega_r\meavsqu\dend}{}
For $\meavsqu =1$ we find
\bigformularray{D'(k,q,r)&=&\frac{8}{\pi k}\int_0^\infty \sin(\lambda k)\sin(\lambda q)\sin(\lambda r)\frac{\dif\lambda}{\lambda}\nonumber\\&=&-\frac{1}{k}\left( {\sgn ( r-q-k) -\sgn ( r-q+k ) -\sgn ( r+q-k) +1}\right)\dend}{}
\section{Relations Involving Bessel Functions of Integer and Fractional Order\label{app:bessel_functions}}

For the reader's convenience we collect some facts about Bessel functions of the first kind $J_n$ and spherical Bessel functions of the first kind $j_n$. They are related by
\bigformula{j_n(z)=\sqrt{\frac{\pi}{2z}}J_{n+1/2}(z)\dend}{bessel_functions_spherical_bessel_functions}

In the main text we employ the following integral of Bessel functions from \cite{Gradshteyn:2000}:
\bigformula{\int_0^\pi \cos\left(\beta\cos\theta\right) J_0\left(\alpha\sin\theta\right)\sin\theta \dif\theta =2\frac{\sin\left(\sqrt{\alpha^2+\beta^2}\right)}{\sqrt{\alpha^2+\beta^2}}\kend}{formula_cos_bessel_sin}
and for the product of two Bessel functions:
\bigformula{J_0\left(\alpha\right)J_0\left(\beta\right)=\frac{1}{\pi}\int_0^\pi J_0\left(\sqrt{\alpha^2+\beta^2-2\alpha\beta\cos\left(x\right)}\right)\dif x\dend}{formula_prod_two_bessel}

Using Rayleigh's formula, the spherical Bessel functions can be computed iteratively from the $\mathrm{sinc}$ function:
\bigformula{\sbessel{n}{z}=z^n\left(-\frac{1}{z}\frac{\dif}{\dif z}\right)^n\frac{\sin(z)}{z}\kend}{rayleighs_formula}

They satisfy the closure relation \cite{Watson:1966}:
\bigformula{\frac{2z^2}{\pi}\int_0^\infty \lambda^2j_n\left(\lambda z\right)j_n\left(\lambda z'\right)\dif\lambda=\delta\left(z-z'\right)\dend}{spherical_bessel_closure_relation}

Several authors have derived expressions or algorithms for the computation of integrals involving products of three spherical Bessel functions \cite{jackson:1972,Anni:1974,Elbaz:1974}
\bigformula{
\intIthree{l_1}{l_2}{l_3}{k}{p}{q}=\int_0^\infty\lambda^2\sbessel{l_1}{k\lambda}\sbessel{l_2}{p\lambda}\sbessel{l_3}{q\lambda}\diff{\lambda}\dend
}{}

Here we cite the explicit result, found by Mehrem et al. \cite{Mehrem:1991} by relating $\intIthree{l_1}{l_2}{l_3}{k}{p}{q}$ to known integrals of three spherical harmonics:
\bigformularray{
\intIthree{l_1}{l_2}{l_3}{k}{p}{q} & = &\nonumber\\ && \hspace{-20mm}\frac{\pi \heaviside{q-\abs{k-p}}\heaviside{k+p-q} i^{l_1+l_2-l_3}}{4kpq}\sqrt{2l_3+1}(k/q)^{l_3}\wignerthreej{l_1}{l_2}{l_3}{0}{0}{0}^{-1}\nonumber\\
& &\hspace{-15mm} \times \sum_{n=0}^{l_3}\left(\begin{matrix}2l_3 2n\end{matrix}\right)^{1/2}(p/k)^n\sum_{l=\abs{l_1-(l_3-n)}}^{l_1+l_3-n}(2l+1) \nonumber\\ & & \hspace{-15mm} \times \wignerthreej{l_1}{l_3-n}{l}{0}{0}{0}\wignerthreej{l_2}{n}{l}{0}{0}{0}\wignersixj{l_1}{l_2}{l_3}{n}{l_3-n}{l}P_l\left(\frac{k^2+p^2-q^2}{2kp}\right)\kend\nonumber\\}{intIthree_result}
where $P_l$ denotes the Legendre polynomials with explicit representation \cite{Abramowitz:1970}
\bigformula{
P_l(x)=\sum_{i=0}^{\floorbr{\frac{l}{2}}}\frac{(-1)^i(2l-2i)!}{2^li!(l-i)!(l-2i)!}\,x^{l-2i}
}{legendre_poly_explicit}

Wigner's 3j symbol is related to the Clebsch-Gordan coefficients by \cite{Edmonds:1954}
\begin{equation}
\begin{pmatrix}
  j_1 & j_2 & j_3\\
  m_1 & m_2 & m_3
\end{pmatrix}
\equiv \frac{(-1)^{j_1-j_2-m_3}}{\sqrt{2j_3+1}} \langle j_1 m_1 j_2 m_2 | j_3 \, {-m_3} \rangle
\end{equation}
Wigner's 3j symbols are equal to zero, unless $m_1+m_2+m_3=0$, $\abs{m_i}\leq j_i$ and $\abs{j_1-j_2}\leq j_3\leq j_1+j_2$

Wigner's 6j symbols are related to Racah's W-coefficients by
\begin{equation}
  \begin{Bmatrix}
    j_1 & j_2 & j_3\\
    j_4 & j_5 & j_6
  \end{Bmatrix}
   = (-1)^{j_1+j_2+j_4+j_5}W(j_1j_2j_5j_4;j_3j_6).
\end{equation}

In the main text, \eqn\eqnref{sbessel_expansion}, we use an expansion into spherical bessel functions \cite{Gonzalez:2000,Bezubik:2004}. We start with the expression (given in \cite{Bezubik:2004}):
\bigformula{\int_{S^{d-1}}e^{i(\vecx|\vecetaunit)}P^\indl (\vecetaunit)\dif \sigma_{\vecetaunit}=\left(\frac{i}{2}\right)^\indl\sum_{\indk =0}^{\floorbr{\frac{\indl}{2}}}\frac{(-1)^k\Gamma (d/2)}{\indk !\,\Gamma (\indl -\indk +d/2)}\tilde{j}_{\indl-\indk+d/2-1}(\vecxabs)(\laplace^\indk P^\indl)(\vecx)\kend}{plane_wave_expansion}
where $\vecx ,\vecetaunit\in \mathbb{R}^d$ with $(\vecetaunit|\vecetaunit)=1$, $\tilde{j}_\nu (z)=\Gamma (\nu +1)\left(\frac{2}{z}\right)^\nu J_\nu (z)$ and the integration is over the $(d-1)$-sphere $S^{d-1}$. $P^\indl$ is a homogeneous polynomial of degree $\indl$ on $\mathbb{R}^d$ and $(.|,)$ denotes the inner product. For $d=3$, we find
\bigformula{\int e^{i\vecx\vecetaunit}P^\indl (\vecetaunit)\dif\Omega_{\vecetaunit} =\left(\frac{i}{2}\right)^\indl\sum_{\indk =0}^{\floorbr{\frac{\indl}{2}}}\frac{(-1)^\indk}{\indk !}\left(\frac{2}{\vecxabs}\right)^{\indl-\indk}\sbessel{\indl-\indk}{\vecxabs}(\laplace^\indk P^\indl)(\vecx)\dend}{plane_wave_expansion_1}
Choosing $P^\indl (\veceta)=(\veckunit\vecetaunit)^\indl$ we get with $(\laplace^\indk P^\indl)(\vecx)={\indl !}/{(\indl -2\indk)!}\,\cdot(\veckunit\vecx)^{\indl-2\indk}$:
\bigformula{\int e^{i\vecx\vecetaunit} (\veckunit\vecetaunit)^\indl \dif\Omega_{\vecetaunit} =i^\indl\sum_{\indk =0}^{\floorbr{\frac{\indl}{2}}}\frac{(-1)^\indk \indl !}{\indk !\,(\indl-2\indk)!}\frac{\sbessel{\indl-\indk}{\vecxabs}}{\left(2\vecxabs\right)^{\indk}}(\veckunit\vecxunit)^{\indl-2\indk}\dend}{plane_wave_expansion_2}
Substituting $\vecx\rightarrow\pm\bvl\vecetaabs$ reproduces \eqn\eqnref{sbessel_expansion}.

  \renewcommand{\theequation}{B-\arabic{equation}}
  % redefine the command that creates the equation no.
  \setcounter{equation}{0}  % reset counter 

\section{The Integrals \texorpdfstring{\mathversion{bold}${\Dnm}$\mathversion{normal}}{Dnm}\label{app:functions_Dnm}}
\newcommand{\TILDEQ}[1]{{#1}^{q\rightarrow -q}}
\newcommand{\TILDER}[1]{{#1}^{r\rightarrow -r}}
\newcommand{\TILDEQR}[1]{{#1}^{q,r\rightarrow -q,-r}}
\newcommand{\alternativeform}[1]{}

\newcommand{\Dnmtabular}[8]{\bigformula{\vspace{-10mm}\Dnmind{#1}{#2}(k,p,q,r):}{D_#1_#2_parameters}\vspace{-9mm}\footnotesize\bigformularray{\salign\termoverall=#3\kend\,\,\,\termremain=#7\kend\nonumber\\ \salign\termpolypref{1}=#4\kend \nonumber \alternativeform{\\ \salign\termpolypref{2}=#5\kend\nonumber\\ \salign\termpolypref{3}=#6\kend\nonumber}}{D_#1_#2_result} \vspace{-9mm}\bigformularray{#8\nonumber}{} \normalsize}
\newcommand{\alternativepolypref}[2]{#1}

The functions $\Dnm (k,p,q,r)$ can all be written in the form 
\bigformularray{\Dnmind{n}{m}(k,p,q,r)=\termoverall\frac{\termprefac{}}{k^{n+m+1}q^n r^m}\big( \termpolypref{1}R_{{1}}+\termpolypref{2}R_{{2}}+\termpolypref{3}R_{{3}}+C\big)\dend}{Dnm_compact_notation_app}
In the following we list the coefficients $\termoverall$, $\termpolypref{1}$ and $\termremain$, which themselves depend on the momenta, for $\Dnm$ with $n+m\leq 5$ and $n\leq m$ ($\Dnmind{n}{m}$ with $n > m$ can be derived from $\Dnmind{m}{n}$ by interchanging $q$ and $r$). The expressions $\termpolypref{2}$ ($\termpolypref{3}$) are found by substituting in $\termpolypref{1}$ the term $c_1$ by $c_2$ ($c_3$) and $f_i$ by $\TILDEQ{f_i}$ ($\TILDER{f_i}$) for all $i$.

\Dnmtabular{0}{0}{1/2}{-1}{-1}{-1}{2\,k}{}
\vspace{-10mm}
\Dnmtabular{0}{1}{-1/12}{f_{{2}}c_{{1}}-{c_{{1}}}^{2}+f_{{1}}}{\alternativepolypref{\termpolypref{1}^{c_1\rightarrow c_2}}{f_{{2}}c_{{2}}-{c_{{2}}}^{2}+f_{{1}}}}{f_{{3}}c_{{3}}-{c_{{3}}}^{2}-f_{{1}}}{-4\,{k}^{3}}{[\salign f_{{1}}=6\,kr,f_{{2}}=3\,k-3\,r\alternativeform{,f_{{3}}=\TILDER{f_{{2}}}}]}
\Dnmtabular{0}{2}{{\frac {1}{120}}}{f_{{2}}c_{{1}}+f_{{3}}{c_{{1}}}^{2}+f_{{4}}{c_{{1}}}^{3}-3\,{c_{{1}}}^{4}+f_{{1}}}{\alternativepolypref{\termpolypref{1}^{c_1\rightarrow c_2}}{f_{{2}}c_{{2}}+f_{{3}}{c_{{2}}}^{2}+f_{{4}}{c_{{2}}}^{3}-3\,{c_{{2}}}^{4}+f_{{1}}}}{f_{{5}}c_{{3}}+f_{{6}}{c_{{3}}}^{2}+f_{{7}}{c_{{3}}}^{3}-3\,{c_{{3}}}^{4}+f_{{1}}}{8\,{k}^{3} \big( 2\,{k}^{2}+5\,{r}^{2} \big) }{[\salign f_{{1}}=-60\,{k}^{2}{r}^{2},f_{{2}}=-60\,kr \big(k-r\big) ,f_{{3}}=-20\,{r}^{2}-20\,{k}^{2}+60\,kr,f_{{4}}=-15\,r+15\,k\alternativeform{,\nonumber\\ \salign f_{{5}}=\TILDER{f_{{2}}},f_{{6}}=\TILDER{f_{{3}}},f_{{7}}=\TILDER{f_{{4}}} } ]}
\Dnmtabular{0}{3}{-{\frac {1}{560}}}{f_{{2}}c_{{1}}+f_{{3}}{c_{{1}}}^{2}+f_{{4}}{c_{{1}}}^{3}+f_{{5}}{c_{{1}}}^{4}+f_{{6}}{c_{{1}}}^{5}-5\,{c_{{1}}}^{6}+f_{{1}}}{\alternativepolypref{\termpolypref{1}^{c_1\rightarrow c_2}}{f_{{2}}c_{{2}}+f_{{3}}{c_{{2}}}^{2}+f_{{4}}{c_{{2}}}^{3}+f_{{5}}{c_{{2}}}^{4}+f_{{6}}{c_{{2}}}^{5}-5\,{c_{{2}}}^{6}+f_{{1}}}}{f_{{7}}c_{{3}}+f_{{8}}{c_{{3}}}^{2}+f_{{9}}{c_{{3}}}^{3}+f_{{10}}{c_{{3}}}^{4}+f_{{11}}{c_{{3}}}^{5}-5\,{c_{{3}}}^{6}-f_{{1}}}{-16\,{k}^{5} \big( 2\,{k}^{2}+7\,{r}^{2} \big) }{[\salign f_{{1}}=280\,{r}^{3}{k}^{3},f_{{2}}=420\,{k}^{2}{r}^{2} \big(k-r\big) ,f_{{3}}=140\,kr \big( 2\,k-r \big)  \big( k-2\,r \big) ,f_{{4}}=70\, \big(k-r\big) \nonumber\\ \salign\times \big( {r}^{2}-5\,kr+{k}^{2} \big) ,f_{{5}}=-42\, \big( 2\,k-r \big)  \big( k-2\,r \big) ,f_{{6}}=-35\,r+35\,k\alternativeform{,f_{{7}}=\TILDER{f_{{2}}},f_{{8}}=\TILDER{f_{{3}}},\nonumber\\ \salign f_{{9}}=\TILDER{f_{{4}}},f_{{10}}=\TILDER{f_{{5}}},f_{{11}}=\TILDER{f_{{6}}} } ]}
\Dnmtabular{0}{4}{{\frac {1}{10080}}}{f_{{2}}c_{{1}}+f_{{3}}{c_{{1}}}^{2}+f_{{4}}{c_{{1}}}^{3}+f_{{5}}{c_{{1}}}^{4}+f_{{6}}{c_{{1}}}^{5}+f_{{7}}{c_{{1}}}^{6}+f_{{8}}{c_{{1}}}^{7}-35\,{c_{{1}}}^{8}+f_{{1}}}{\alternativepolypref{\termpolypref{1}^{c_1\rightarrow c_2}}{f_{{2}}c_{{2}}+f_{{3}}{c_{{2}}}^{2}+f_{{4}}{c_{{2}}}^{3}+f_{{5}}{c_{{2}}}^{4}+f_{{6}}{c_{{2}}}^{5}+f_{{7}}{c_{{2}}}^{6}+f_{{8}}{c_{{2}}}^{7}-35\,{c_{{2}}}^{8}+f_{{1}}}}{f_{{9}}c_{{3}}+f_{{10}}{c_{{3}}}^{2}+f_{{11}}{c_{{3}}}^{3}+f_{{12}}{c_{{3}}}^{4}+f_{{13}}{c_{{3}}}^{5}+f_{{14}}{c_{{3}}}^{6}+f_{{15}}{c_{{3}}}^{7}-35\,{c_{{3}}}^{8}+f_{{1}}}{32\,{k}^{5} \big( 36\,{k}^{2}{r}^{2}+63\,{r}^{4}+8\,{k}^{4} \big) }{[\salign f_{{1}}=-5040\,{k}^{4}{r}^{4},f_{{2}}=-10080\,{k}^{3}{r}^{3} \big(k-r\big) ,f_{{3}}=-3360\,{k}^{2}{r}^{2} \big( 3\,{r}^{2}+3\,{k}^{2}-7\,kr \big) ,\nonumber\\ \salign f_{{4}}=-2520\,kr \big(k-r\big)  \big( 2\,{r}^{2}-7\,kr+2\,{k}^{2} \big) ,f_{{5}}=10080\,{k}^{3}r-1008\,{r}^{4}+10080\,k{r}^{3}-1008\,{k}^{4}\nonumber\\ \salign -19656\,{k}^{2}{r}^{2},f_{{6}}=840\, \big(k-r\big)  \big( 2\,{r}^{2}-7\,kr+2\,{k}^{2} \big) ,f_{{7}}=2520\,kr-1080\,{k}^{2}-1080\,{r}^{2},\nonumber\\ \salign f_{{8}}=-315\,r+315\,k\alternativeform{,f_{{9}}=\TILDER{f_{{2}}},f_{{10}}=\TILDER{f_{{3}}},f_{{11}}=\TILDER{f_{{4}}},f_{{12}}=\TILDER{f_{{5}}},f_{{13}}=\TILDER{f_{{6}}},\nonumber\\ \salign f_{{14}}=\TILDER{f_{{7}}},f_{{15}}=\TILDER{f_{{8}}} } ]}
\Dnmtabular{0}{5}{-{\frac {1}{44352}}}{f_{{2}}c_{{1}}+f_{{3}}{c_{{1}}}^{2}+f_{{4}}{c_{{1}}}^{3}+f_{{5}}{c_{{1}}}^{4}+f_{{6}}{c_{{1}}}^{5}+f_{{7}}{c_{{1}}}^{6}+f_{{8}}{c_{{1}}}^{7}+f_{{9}}{c_{{1}}}^{8}+f_{{10}}{c_{{1}}}^{9}-63\,{c_{{1}}}^{10}+f_{{1}}}{\alternativepolypref{\termpolypref{1}^{c_1\rightarrow c_2}}{f_{{2}}c_{{2}}+f_{{3}}{c_{{2}}}^{2}+f_{{4}}{c_{{2}}}^{3}+f_{{5}}{c_{{2}}}^{4}+f_{{6}}{c_{{2}}}^{5}+f_{{7}}{c_{{2}}}^{6}+f_{{8}}{c_{{2}}}^{7}+f_{{9}}{c_{{2}}}^{8}+f_{{10}}{c_{{2}}}^{9}-63\,{c_{{2}}}^{10}+f_{{1}}}}{f_{{11}}c_{{3}}+f_{{12}}{c_{{3}}}^{2}+f_{{13}}{c_{{3}}}^{3}+f_{{14}}{c_{{3}}}^{4}+f_{{15}}{c_{{3}}}^{5}+f_{{16}}{c_{{3}}}^{6}+f_{{17}}{c_{{3}}}^{7}+f_{{18}}{c_{{3}}}^{8}+f_{{19}}{c_{{3}}}^{9}-63\,{c_{{3}}}^{10}-f_{{1}}}{-64\,{k}^{7} \big( 44\,{k}^{2}{r}^{2}+8\,{k}^{4}+99\,{r}^{4} \big) }{[\salign f_{{1}}=22176\,{r}^{5}{k}^{5},f_{{2}}=55440\,{k}^{4}{r}^{4} \big(k-r\big) ,f_{{3}}=18480\,{k}^{3}{r}^{3} \big( 4\,{r}^{2}-9\,kr+4\,{k}^{2} \big) ,f_{{4}}=55440\,{k}^{2}{r}^{2}\nonumber\\ \salign\times \big(k-r\big)  \big( {r}^{2}-3\,kr+{k}^{2} \big) ,f_{{5}}=11088\,kr \big( -14\,{k}^{3}r+2\,{r}^{4}-14\,k{r}^{3}+25\,{k}^{2}{r}^{2}+2\,{k}^{4} \big) ,\nonumber\\ \salign f_{{6}}=1848\, \big(k-r\big)  \big( 2\,{r}^{4}-28\,k{r}^{3}+67\,{k}^{2}{r}^{2}-28\,{k}^{3}r+2\,{k}^{4} \big) ,f_{{7}}=-99000\,{k}^{2}{r}^{2}-7920\,{k}^{4}\nonumber\\ \salign -7920\,{r}^{4}+55440\,k{r}^{3}+55440\,{k}^{3}r,f_{{8}}=6930\, \big(k-r\big)  \big( {r}^{2}-3\,kr+{k}^{2} \big) ,f_{{9}}=-3080\,{k}^{2}-3080\,{r}^{2}\nonumber\\ \salign +6930\,kr,f_{{10}}=-693\,r+693\,k\alternativeform{,f_{{11}}=\TILDER{f_{{2}}},f_{{12}}=\TILDER{f_{{3}}},f_{{13}}=\TILDER{f_{{4}}},f_{{14}}=\TILDER{f_{{5}}},\nonumber\\ \salign f_{{15}}=\TILDER{f_{{6}}},f_{{16}}=\TILDER{f_{{7}}},f_{{17}}=\TILDER{f_{{8}}},f_{{18}}=\TILDER{f_{{9}}},f_{{19}}=\TILDER{f_{{10}}} }
 ]}
\Dnmtabular{1}{1}{{\frac {1}{120}}}{f_{{2}}c_{{1}}+f_{{3}}{c_{{1}}}^{2}+f_{{4}}{c_{{1}}}^{3}-{c_{{1}}}^{4}+f_{{1}}}{f_{{5}}c_{{2}}+f_{{6}}{c_{{2}}}^{2}+f_{{7}}{c_{{2}}}^{3}-{c_{{2}}}^{4}-f_{{1}}}{f_{{8}}c_{{3}}+f_{{9}}{c_{{3}}}^{2}+f_{{10}}{c_{{3}}}^{3}-{c_{{3}}}^{4}-f_{{1}}}{4\,{k}^{3} \big( 3\,{k}^{2}+5\,{p}^{2}-5\,{q}^{2}-5\,{r}^{2} \big) }{[\salign f_{{1}}=-60\,{k}^{2}qr,f_{{2}}=-30\,k \big( -2\,qr+kq+kr \big) ,f_{{3}}=20\,kq+20\,kr-20\,qr-10\,{k}^{2},\nonumber\\ \salign f_{{4}}=-5\,q-5\,r+5\,k\alternativeform{,f_{{5}}=\TILDEQ{f_{{2}}},f_{{6}}=\TILDEQ{f_{{3}}},f_{{7}}=\TILDEQ{f_{{4}}},f_{{8}}=\TILDER{f_{{2}}},\nonumber\\ \salign f_{{9}}=\TILDER{f_{{3}}},f_{{10}}=\TILDER{f_{{4}}} } ]}
\Dnmtabular{1}{2}{-{\frac {1}{1680}}}{f_{{2}}c_{{1}}+f_{{3}}{c_{{1}}}^{2}+f_{{4}}{c_{{1}}}^{3}+f_{{5}}{c_{{1}}}^{4}+f_{{6}}{c_{{1}}}^{5}-3\,{c_{{1}}}^{6}+f_{{1}}}{f_{{7}}c_{{2}}+f_{{8}}{c_{{2}}}^{2}+f_{{9}}{c_{{2}}}^{3}+f_{{10}}{c_{{2}}}^{4}+f_{{11}}{c_{{2}}}^{5}-3\,{c_{{2}}}^{6}-f_{{1}}}{f_{{12}}c_{{3}}+f_{{13}}{c_{{3}}}^{2}+f_{{14}}{c_{{3}}}^{3}+f_{{15}}{c_{{3}}}^{4}+f_{{16}}{c_{{3}}}^{5}-3\,{c_{{3}}}^{6}+f_{{1}}}{-16\,{k}^{5} \big( -14\,{q}^{2}+14\,{p}^{2}+7\,{r}^{2}+4\,{k}^{2} \big) }{[\salign f_{{1}}=840\,q{r}^{2}{k}^{3},f_{{2}}=420\,{k}^{2}r \big( kr-3\,qr+2\,kq \big) ,f_{{3}}=140\,k \big( 2\,r{k}^{2}+6\,{r}^{2}q-3\,{r}^{2}k+2\,q{k}^{2}-8\,rqk \big) ,\nonumber\\ \salign f_{{4}}=-280\,q{k}^{2}+630\,rqk+70\,{k}^{3}-280\,r{k}^{2}-210\,{r}^{2}q+210\,{r}^{2}k,f_{{5}}=126\,kr-126\,qr-42\,{r}^{2}\nonumber\\ \salign -56\,{k}^{2}+126\,kq,f_{{6}}=21\,k-21\,r-21\,q\alternativeform{,f_{{7}}=\TILDEQ{f_{{2}}},f_{{8}}=\TILDEQ{f_{{3}}},f_{{9}}=\TILDEQ{f_{{4}}},f_{{10}}=\TILDEQ{f_{{5}}},\nonumber\\ \salign f_{{11}}=\TILDEQ{f_{{6}}},f_{{12}}=\TILDER{f_{{2}}},f_{{13}}=\TILDER{f_{{3}}},f_{{14}}=\TILDER{f_{{4}}},f_{{15}}=\TILDER{f_{{5}}},f_{{16}}=\TILDER{f_{{6}}} } ]}
\Dnmtabular{1}{3}{{\frac {1}{10080}}}{f_{{2}}c_{{1}}+f_{{3}}{c_{{1}}}^{2}+f_{{4}}{c_{{1}}}^{3}+f_{{5}}{c_{{1}}}^{4}+f_{{6}}{c_{{1}}}^{5}+f_{{7}}{c_{{1}}}^{6}+f_{{8}}{c_{{1}}}^{7}-5\,{c_{{1}}}^{8}+f_{{1}}}{f_{{9}}c_{{2}}+f_{{10}}{c_{{2}}}^{2}+f_{{11}}{c_{{2}}}^{3}+f_{{12}}{c_{{2}}}^{4}+f_{{13}}{c_{{2}}}^{5}+f_{{14}}{c_{{2}}}^{6}+f_{{15}}{c_{{2}}}^{7}-5\,{c_{{2}}}^{8}-f_{{1}}}{f_{{16}}c_{{3}}+f_{{17}}{c_{{3}}}^{2}+f_{{18}}{c_{{3}}}^{3}+f_{{19}}{c_{{3}}}^{4}+f_{{20}}{c_{{3}}}^{5}+f_{{21}}{c_{{3}}}^{6}+f_{{22}}{c_{{3}}}^{7}-5\,{c_{{3}}}^{8}-f_{{1}}}{16\,{k}^{5} \big( 54\,{k}^{2}{p}^{2}-63\,{q}^{2}{r}^{2}-54\,{q}^{2}{k}^{2}-63\,{r}^{4}+27\,{k}^{2}{r}^{2}+63\,{p}^{2}{r}^{2}+10\,{k}^{4} \big) }{[\salign f_{{1}}=-5040\,{r}^{3}{k}^{4}q,f_{{2}}=-2520\,{k}^{3}{r}^{2} \big( kr-4\,qr+3\,kq \big) ,f_{{3}}=-840\,{k}^{2}r \big( -4\,{r}^{2}k+12\,{r}^{2}q+3\,r{k}^{2}\nonumber\\ \salign -18\,rqk+6\,q{k}^{2} \big) ,f_{{4}}=-1260\,k \big( -4\,{r}^{3}q+11\,kq{r}^{2}-7\,{k}^{2}qr+q{k}^{3}-3\,{k}^{2}{r}^{2}+{k}^{3}r+2\,k{r}^{3} \big) ,\nonumber\\ \salign f_{{5}}=6048\,kq{r}^{2}+1764\,{k}^{3}r+1008\,k{r}^{3}+1764\,q{k}^{3}-1008\,{r}^{3}q-252\,{k}^{4}-6804\,{k}^{2}qr-2772\,{k}^{2}{r}^{2},\nonumber\\ \salign f_{{6}}=1008\,{r}^{2}k+2520\,rqk-1008\,{r}^{2}q-1134\,q{k}^{2}-1134\,r{k}^{2}-168\,{r}^{3}+294\,{k}^{3},f_{{7}}=360\,kr\nonumber\\ \salign -360\,qr-144\,{r}^{2}+360\,kq-162\,{k}^{2},f_{{8}}=45\,k-45\,q-45\,r\alternativeform{,f_{{9}}=\TILDEQ{f_{{2}}},f_{{10}}=\TILDEQ{f_{{3}}},\nonumber\\ \salign f_{{11}}=\TILDEQ{f_{{4}}},f_{{12}}=\TILDEQ{f_{{5}}},f_{{13}}=\TILDEQ{f_{{6}}},f_{{14}}=\TILDEQ{f_{{7}}},f_{{15}}=\TILDEQ{f_{{8}}},f_{{16}}=\TILDER{f_{{2}}},\nonumber\\ \salign f_{{17}}=\TILDER{f_{{3}}},f_{{18}}=\TILDER{f_{{4}}},f_{{19}}=\TILDER{f_{{5}}},f_{{20}}=\TILDER{f_{{6}}},f_{{21}}=\TILDER{f_{{7}}},f_{{22}}=\TILDER{f_{{8}}} } ]}
\Dnmtabular{1}{4}{-{\frac {1}{221760}}}{f_{{2}}c_{{1}}+f_{{3}}{c_{{1}}}^{2}+f_{{4}}{c_{{1}}}^{3}+f_{{5}}{c_{{1}}}^{4}+f_{{6}}{c_{{1}}}^{5}+f_{{7}}{c_{{1}}}^{6}+f_{{8}}{c_{{1}}}^{7}+f_{{9}}{c_{{1}}}^{8}+f_{{10}}{c_{{1}}}^{9}-35\,{c_{{1}}}^{10}+f_{{1}}}{f_{{11}}c_{{2}}+f_{{12}}{c_{{2}}}^{2}+f_{{13}}{c_{{2}}}^{3}+f_{{14}}{c_{{2}}}^{4}+f_{{15}}{c_{{2}}}^{5}+f_{{16}}{c_{{2}}}^{6}+f_{{17}}{c_{{2}}}^{7}+f_{{18}}{c_{{2}}}^{8}+f_{{19}}{c_{{2}}}^{9}-35\,{c_{{2}}}^{10}-f_{{1}}}{f_{{20}}c_{{3}}+f_{{21}}{c_{{3}}}^{2}+f_{{22}}{c_{{3}}}^{3}+f_{{23}}{c_{{3}}}^{4}+f_{{24}}{c_{{3}}}^{5}+f_{{25}}{c_{{3}}}^{6}+f_{{26}}{c_{{3}}}^{7}+f_{{27}}{c_{{3}}}^{8}+f_{{28}}{c_{{3}}}^{9}-35\,{c_{{3}}}^{10}+f_{{1}}}{-64\,{k}^{7} \big( 99\,{r}^{4}+88\,{k}^{2}{r}^{2}+396\,{p}^{2}{r}^{2}-396\,{q}^{2}{r}^{2}+176\,{k}^{2}{p}^{2}-176\,{q}^{2}{k}^{2}+24\,{k}^{4} \big) }{[\salign f_{{1}}=110880\,{r}^{4}q{k}^{5},f_{{2}}=55440\,{k}^{4}{r}^{3} \big( -5\,qr+kr+4\,kq \big) ,f_{{3}}=18480\,{k}^{3}{r}^{2} \big( 20\,{r}^{2}q-5\,{r}^{2}k\nonumber\\ \salign -32\,rqk+4\,r{k}^{2}+12\,q{k}^{2} \big) ,f_{{4}}=18480\,{k}^{2}r \big( 5\,k{r}^{3}-15\,{r}^{3}q-8\,{k}^{2}{r}^{2}+41\,kq{r}^{2}-30\,{k}^{2}qr\nonumber\\ \salign +3\,{k}^{3}r+6\,q{k}^{3} \big) ,f_{{5}}=3696\,k \big( -15\,{r}^{4}k+30\,{r}^{4}q-66\,{k}^{3}qr+6\,q{k}^{4}-141\,k{r}^{3}q+6\,{k}^{4}r\nonumber\\ \salign -30\,{k}^{3}{r}^{2}+171\,{k}^{2}q{r}^{2}+41\,{k}^{2}{r}^{3} \big) ,f_{{6}}=18480\,{r}^{4}k-40656\,{k}^{4}r+240240\,{k}^{3}qr+105336\,{k}^{3}{r}^{2}\nonumber\\ \salign +184800\,k{r}^{3}q-18480\,{r}^{4}q-86856\,{k}^{2}{r}^{3}+3696\,{k}^{5}-40656\,q{k}^{4}-382536\,{k}^{2}q{r}^{2},f_{{7}}=-2640\,{r}^{4}\nonumber\\ \salign -126720\,{k}^{2}qr-26400\,{r}^{3}q+118800\,kq{r}^{2}-5808\,{k}^{4}+34320\,{k}^{3}r+34320\,q{k}^{3}+26400\,k{r}^{3}\nonumber\\ \salign -54648\,{k}^{2}{r}^{2},f_{{8}}=14850\,{r}^{2}k-14850\,{r}^{2}q-15840\,r{k}^{2}+34650\,rqk-15840\,q{k}^{2}+4290\,{k}^{3}\nonumber\\ \salign -3300\,{r}^{3},f_{{9}}=-1760\,{k}^{2}+3850\,kq-1650\,{r}^{2}-3850\,qr+3850\,kr,f_{{10}}=-385\,r-385\,q\nonumber\\ \salign +385\,k\alternativeform{,f_{{11}}=\TILDEQ{f_{{2}}},f_{{12}}=\TILDEQ{f_{{3}}},f_{{13}}=\TILDEQ{f_{{4}}},f_{{14}}=\TILDEQ{f_{{5}}},f_{{15}}=\TILDEQ{f_{{6}}},\nonumber\\ \salign f_{{16}}=\TILDEQ{f_{{7}}},f_{{17}}=\TILDEQ{f_{{8}}},f_{{18}}=\TILDEQ{f_{{9}}},f_{{19}}=\TILDEQ{f_{{10}}},f_{{20}}=\TILDER{f_{{2}}},f_{{21}}=\TILDER{f_{{3}}},\nonumber\\ \salign f_{{22}}=\TILDER{f_{{4}}},f_{{23}}=\TILDER{f_{{5}}},f_{{24}}=\TILDER{f_{{6}}},f_{{25}}=\TILDER{f_{{7}}},f_{{26}}=\TILDER{f_{{8}}},f_{{27}}=\TILDER{f_{{9}}},\nonumber\\ \salign f_{{28}}=\TILDER{f_{{10}}} } ]}
\Dnmtabular{2}{2}{{\frac {1}{3360}}}{f_{{2}}c_{{1}}+f_{{3}}{c_{{1}}}^{2}+f_{{4}}{c_{{1}}}^{3}+f_{{5}}{c_{{1}}}^{4}+f_{{6}}{c_{{1}}}^{5}+f_{{7}}{c_{{1}}}^{6}+f_{{8}}{c_{{1}}}^{7}-{c_{{1}}}^{8}+f_{{1}}}{f_{{9}}c_{{2}}+f_{{10}}{c_{{2}}}^{2}+f_{{11}}{c_{{2}}}^{3}+f_{{12}}{c_{{2}}}^{4}+f_{{13}}{c_{{2}}}^{5}+f_{{14}}{c_{{2}}}^{6}+f_{{15}}{c_{{2}}}^{7}-{c_{{2}}}^{8}+f_{{1}}}{f_{{16}}c_{{3}}+f_{{17}}{c_{{3}}}^{2}+f_{{18}}{c_{{3}}}^{3}+f_{{19}}{c_{{3}}}^{4}+f_{{20}}{c_{{3}}}^{5}+f_{{21}}{c_{{3}}}^{6}+f_{{22}}{c_{{3}}}^{7}-{c_{{3}}}^{8}+f_{{1}}}{16\,{k}^{5} \big( 28\,{q}^{2}{r}^{2}+7\,{r}^{4}+22\,{k}^{2}{p}^{2}+7\,{p}^{4}+2\,{q}^{2}{k}^{2}+3\,{k}^{4}-14\,{p}^{2}{r}^{2}\nonumber\\ \salign \hspace{33mm} -14\,{p}^{2}{q}^{2}+2\,{k}^{2}{r}^{2}+7\,{q}^{4} \big) }{[\salign f_{{1}}=-1680\,{q}^{2}{k}^{4}{r}^{2},f_{{2}}=-1680\,{k}^{3}qr \big( -2\,qr+kq+kr \big) ,f_{{3}}=-560\,{k}^{2} \big( kr-3\,qr+kq \big) \nonumber\\ \salign\times \big( -2\,qr+kq+kr \big) ,f_{{4}}=-140\,k \big( 12\,qr-5\,kr-5\,kq+2\,{k}^{2} \big)  \big( -qr+kr+kq \big) ,\nonumber\\ \salign f_{{5}}=-336\,{q}^{2}{r}^{2}+1008\,kq{r}^{2}+1008\,k{q}^{2}r-476\,{q}^{2}{k}^{2}+336\,{k}^{3}r-1344\,{k}^{2}qr+336\,q{k}^{3}\nonumber\\ \salign -476\,{k}^{2}{r}^{2}-56\,{k}^{4},f_{{6}}=56\, \big( -q-r+k \big)  \big( 3\,qr-3\,kq+{k}^{2}-3\,kr \big) ,f_{{7}}=-72\,qr-24\,{q}^{2}\nonumber\\ \salign -24\,{r}^{2}-32\,{k}^{2}+72\,kq+72\,kr,f_{{8}}=-9\,q-9\,r+9\,k\alternativeform{,f_{{9}}=\TILDEQ{f_{{2}}},f_{{10}}=\TILDEQ{f_{{3}}},\nonumber\\ \salign f_{{11}}=\TILDEQ{f_{{4}}},f_{{12}}=\TILDEQ{f_{{5}}},f_{{13}}=\TILDEQ{f_{{6}}},f_{{14}}=\TILDEQ{f_{{7}}},f_{{15}}=\TILDEQ{f_{{8}}},f_{{16}}=\TILDER{f_{{2}}},\nonumber\\ \salign f_{{17}}=\TILDER{f_{{3}}},f_{{18}}=\TILDER{f_{{4}}},f_{{19}}=\TILDER{f_{{5}}},f_{{20}}=\TILDER{f_{{6}}},f_{{21}}=\TILDER{f_{{7}}},f_{{22}}=\TILDER{f_{{8}}} } ]}
\Dnmtabular{2}{3}{-{\frac {1}{221760}}}{f_{{2}}c_{{1}}+f_{{3}}{c_{{1}}}^{2}+f_{{4}}{c_{{1}}}^{3}+f_{{5}}{c_{{1}}}^{4}+f_{{6}}{c_{{1}}}^{5}+f_{{7}}{c_{{1}}}^{6}+f_{{8}}{c_{{1}}}^{7}+f_{{9}}{c_{{1}}}^{8}+f_{{10}}{c_{{1}}}^{9}-15\,{c_{{1}}}^{10}+f_{{1}}}{f_{{11}}c_{{2}}+f_{{12}}{c_{{2}}}^{2}+f_{{13}}{c_{{2}}}^{3}+f_{{14}}{c_{{2}}}^{4}+f_{{15}}{c_{{2}}}^{5}+f_{{16}}{c_{{2}}}^{6}+f_{{17}}{c_{{2}}}^{7}+f_{{18}}{c_{{2}}}^{8}+f_{{19}}{c_{{2}}}^{9}-15\,{c_{{2}}}^{10}+f_{{1}}}{f_{{20}}c_{{3}}+f_{{21}}{c_{{3}}}^{2}+f_{{22}}{c_{{3}}}^{3}+f_{{23}}{c_{{3}}}^{4}+f_{{24}}{c_{{3}}}^{5}+f_{{25}}{c_{{3}}}^{6}+f_{{26}}{c_{{3}}}^{7}+f_{{27}}{c_{{3}}}^{8}+f_{{28}}{c_{{3}}}^{9}-15\,{c_{{3}}}^{10}-f_{{1}}}{-32\,{k}^{7} \big( 297\,{p}^{4}+198\,{p}^{2}{r}^{2}-22\,{q}^{2}{k}^{2}+297\,{q}^{4}+66\,{k}^{2}{r}^{2}+462\,{k}^{2}{p}^{2}\nonumber\\ \salign \hspace{40mm} -495\,{r}^{4}+396\,{q}^{2}{r}^{2}+41\,{k}^{4}-594\,{p}^{2}{q}^{2} \big) }{[\salign f_{{1}}=110880\,{q}^{2}{r}^{3}{k}^{5},f_{{2}}=55440\,{k}^{4}q{r}^{2} \big( -5\,qr+3\,kq+2\,kr \big) ,f_{{3}}=18480\,{k}^{3}r \big( 20\,{q}^{2}{r}^{2}+6\,{k}^{2}qr\nonumber\\ \salign +6\,{q}^{2}{k}^{2}-21\,k{q}^{2}r-12\,kq{r}^{2}+2\,{k}^{2}{r}^{2} \big) ,f_{{4}}=27720\,{k}^{2} \big( 2\,{k}^{3}qr+17\,k{q}^{2}{r}^{2}-8\,{k}^{2}q{r}^{2}-10\,{q}^{2}{r}^{3}\nonumber\\ \salign +{q}^{2}{k}^{3}+{k}^{3}{r}^{2}+9\,k{r}^{3}q-2\,{k}^{2}{r}^{3}-8\,{k}^{2}{q}^{2}r \big) ,f_{{5}}=5544\,k \big( 9\,{k}^{2}{r}^{3}-8\,{k}^{3}{r}^{2}+20\,{q}^{2}{r}^{3}+45\,{k}^{2}q{r}^{2}\nonumber\\ \salign +2\,{k}^{4}r-57\,k{q}^{2}{r}^{2}-18\,{k}^{3}qr-29\,k{r}^{3}q+2\,q{k}^{4}-8\,{q}^{2}{k}^{3}+41\,{k}^{2}{q}^{2}r \big) ,f_{{6}}=110880\,k{q}^{2}{r}^{2}\nonumber\\ \salign -152460\,{k}^{2}q{r}^{2}+99792\,{k}^{3}qr+37884\,{q}^{2}{k}^{3}-18480\,{q}^{2}{r}^{3}-130284\,{k}^{2}{q}^{2}r-16632\,{k}^{4}r+41580\,{k}^{3}{r}^{2}\nonumber\\ \salign -26796\,{k}^{2}{r}^{3}+55440\,k{r}^{3}q+1848\,{k}^{5}-16632\,q{k}^{4},f_{{7}}=-21780\,{k}^{2}{r}^{2}-2376\,{k}^{4}-7920\,{r}^{3}q\nonumber\\ \salign +14256\,q{k}^{3}-53856\,{k}^{2}qr+39600\,k{q}^{2}r-18612\,{q}^{2}{k}^{2}+47520\,kq{r}^{2}+7920\,k{r}^{3}-15840\,{q}^{2}{r}^{2}\nonumber\\ \salign +14256\,{k}^{3}r,f_{{8}}=198\, \big( -q-r+k \big)  \big( -25\,kq+25\,qr+9\,{k}^{2}-25\,kr+5\,{r}^{2} \big) ,f_{{9}}=-660\,{r}^{2}\nonumber\\ \salign -748\,{k}^{2}-550\,{q}^{2}+1650\,kr+1650\,kq-1650\,qr,f_{{10}}=-165\,q+165\,k-165\,r\alternativeform{,f_{{11}}=\TILDEQ{f_{{2}}},\nonumber\\ \salign f_{{12}}=\TILDEQ{f_{{3}}},f_{{13}}=\TILDEQ{f_{{4}}},f_{{14}}=\TILDEQ{f_{{5}}},f_{{15}}=\TILDEQ{f_{{6}}},f_{{16}}=\TILDEQ{f_{{7}}},f_{{17}}=\TILDEQ{f_{{8}}},\nonumber\\ \salign f_{{18}}=\TILDEQ{f_{{9}}},f_{{19}}=\TILDEQ{f_{{10}}},f_{{20}}=\TILDER{f_{{2}}},f_{{21}}=\TILDER{f_{{3}}},f_{{22}}=\TILDER{f_{{4}}},f_{{23}}=\TILDER{f_{{5}}},\nonumber\\ \salign f_{{24}}=\TILDER{f_{{6}}},f_{{25}}=\TILDER{f_{{7}}},f_{{26}}=\TILDER{f_{{8}}},f_{{27}}=\TILDER{f_{{9}}},f_{{28}}=\TILDER{f_{{10}}} } ]}

The $\Dnm$'s do not possess any singularities inside the domain of integration, which is an almost essential feature for the numerical solution of the Boltzmann equation. Note, however, that the expressions given here are neither optimised for numerical efficiency nor for stability. Not all of the terms $\termpolypref{i}$ need to be computed in every step since $R_i=\ramp(c_{i})=0$ for $c_i\leq 0$, and quantities such as powers of the momenta, which appear several times, need to be computed only once for all $\Dnm$'s. In an implementation according to the model presented in \sect\secref{a_simple_numerical_model} the $\Dnm$'s need to be computed only once for all matrix elements in the system (with the possible exception of $\termremain$ and the $c_i$ terms including $p$ which is determined by energy conservation).

\end{appendix}

 \def\eprinttmp@#1arXiv:#2 [#3]#4@{
 \ifthenelse{\equal{#3}{x}}{\href{http://arxiv.org/abs/#1}{\texttt{#1}}}{\href{http://arxiv.org/abs/#1}{\texttt{arXiv:#1}}}
}

 \newcommand{\eprint}[1]{\eprinttmp@#1arXiv: [x]@}

\providecommand{\bysame}{\leavevmode\hbox to3em{\hrulefill}\thinspace}

\noindent 

\newpage

\end{document}